\pdfoutput=1
\documentclass[fleqn,10pt]{wlscirep}
\usepackage[utf8]{inputenc}
\usepackage[T1]{fontenc}
\usepackage{amsmath,amsfonts,bm,float,xcolor}
\usepackage{framed}
\usepackage{physics}
\usepackage{mathpazo}
\usepackage{siunitx}
\usepackage{textalpha}
\usepackage{newtxtext,newtxmath}

\newcommand{\rev}[1]{#1}

\title{Synthetic staggered fields switch Néel vectors probed by magnetic resonance}

\author[1,2,*]{Yifang Hao}
\author[1,2]{Zhiyi Che}
\author[1,2]{Jack Brierley}
\author[3,4]{Aakanksha Sud}
\author[3,5,6]{Satoshi Iihama}
\author[7,8]{Rishi Siddani}
\author[1,2]{Chenghao Yang}
\author[1,2,7]{Charlie W. F. Freeman}
\author[9,10]{Troy Dion}
\author[1,2]{Daniel Prestwood}
\author[1,2]{Zekun Xue}
\author[1,2]{Tianyi Wei}
\author[1,7]{Murat Cubukcu}
\author[7]{Craig Barton}
\author[4,5,9]{Shunsuke Fukami}
\author[5,9]{Shigemi Mizukami}
\author[1,2,5,9,10,*]{Hidekazu Kurebayashi}

\affil[1]{London Centre for Nanotechnology, University College London, 17-19 Gordon Street, London, WC1H 0AH, UK}
\affil[2]{Department of Electronic and Electrical Engineering, University College London, 17-19 Gordon Street, London, WC1H 0AH, UK }
\affil[3]{Frontier Research Institute for Interdisciplinary Sciences, Tohoku University, 6-3 Aoba, Sendai 980-8578, Japan}
\affil[4]{Research Institute of Electrical Communication, Tohoku University, 2-1-1 Katahira, Sendai 980-8577, Japan}
\affil[5]{WPI Advanced Institute for Materials Research, Tohoku University, 2-1-1 Katahira, Sendai 980-8577, Japan}
\affil[6]{Department of Materials Physics, Nagoya University, Nagoya 464-8603, Japan}
\affil[7]{National Physical Laboratory, Teddington, TW11 0LW, UK}
\affil[8]{Blackett Laboratory, Imperial College London, Prince Consort Road, London, SW7 2AZ, UK}
\affil[9]{Center for Science and Innovation in Spintronics, Tohoku University, 2-1-1, Katahira, Sendai, 980-8577 Japan}
\affil[10]{Institute for Materials Research, Tohoku University, 2-1-1, Katahira, Sendai, 980-8577 Japan}
\affil[*]{Corresponding author e-mail: uceeyh1@ucl.ac.uk; h.kurebayashi@ucl.ac.uk}

\begin{abstract}
Electrical detection of antiferromagnetic order is a critical challenge for antiferromagnetic spintronics owing to their vanishing macroscopic magnetisation. While magnetic resonance has demonstrated sub-terahertz antiferromagnetic spin pumping and nonlinear magnon dynamics, its application to electrically contacted on-chip devices has room for further exploration. Here we show that magnetic resonance enables direct electrical detection of Néel vector states. A 180$^\circ$ reversal in synthetic antiferromagnets is realised by current-driven effective fields with opposite polarisation acting on each magnetic sublayer, which we term synthetic staggered fields. The switching induces a $\pi$ phase shift of precessional motion in time, producing hysteretic rectified voltages in our phase-sensitive detection scheme. Angular-dependent resonance analysis offers further insight into the Néel vector directions, pointing to a model beyond the macrospin formalism. We address this by defect-incorporated micromagnetic simulations that reveal fragmented domain states after current excitation. \rev{An extracted} current density of \rev{7.4} $\times$ 10$^9$ A/m$^2$ is, to our knowledge, the lowest reported value for the room-temperature Néel vector switching and we show that it scales with the in-plane uniaxial anisotropy both experimentally and numerically. Synthetic staggered fields and the resonance-based methodology provide a research platform for nanoscale characterisation of electrically driven antiferromagnetic textures, guiding the development of high-frequency on-chip spintronic architectures.
\end{abstract}

\begin{document}
\hbadness=10000
\flushbottom
\maketitle

\section*{Main}
\begin{figure}[ht!]
    \centering
    \includegraphics[width=1.0\linewidth]{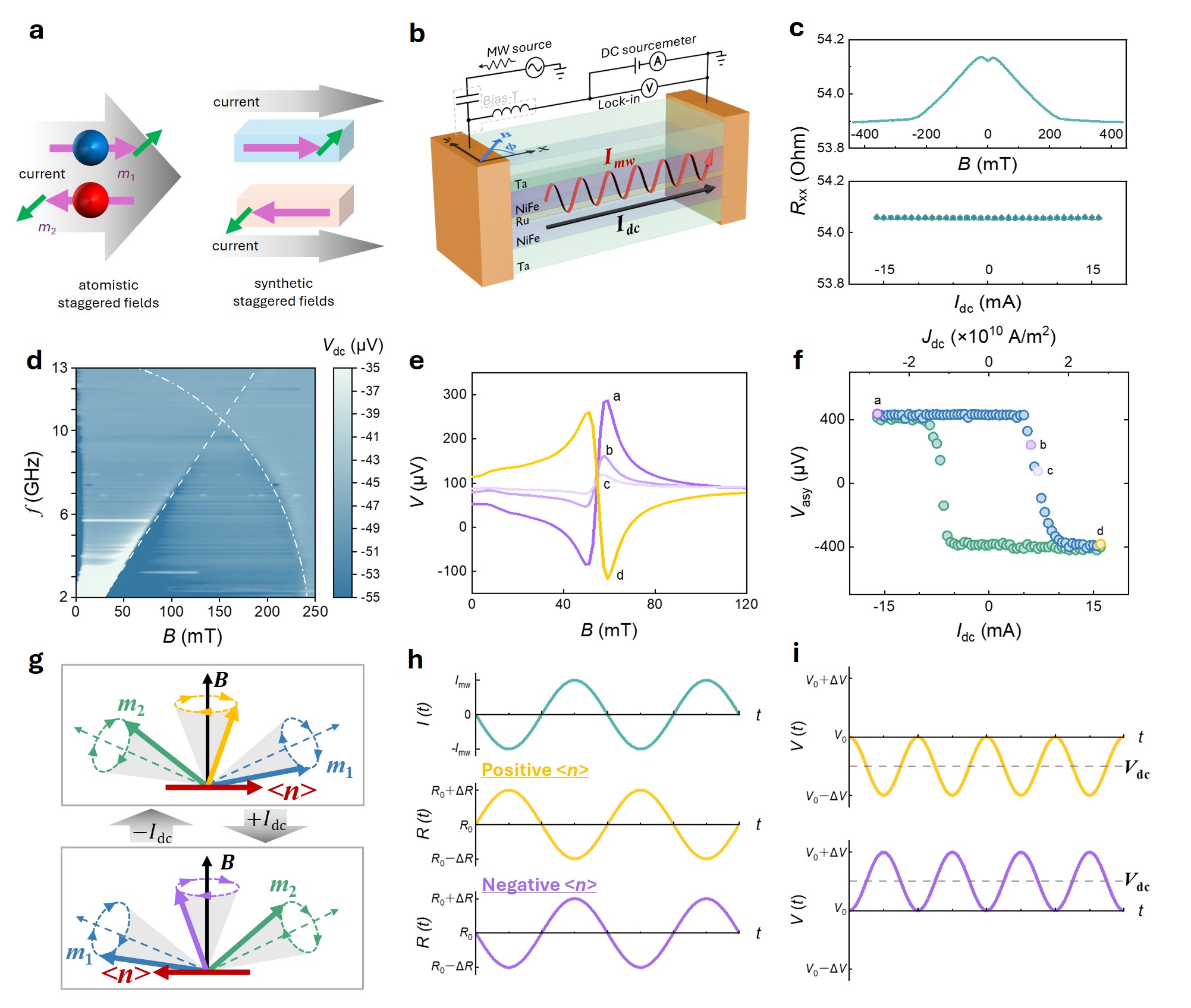}
    \caption{\textbf{Synthetic staggered fields and their use for Néel vector switching measured by magnetic resonance.} \textbf{a}, Schematics of atomistic and synthetic staggered fields. Current-induced fields (green arrows) point oppositely for each sub-lattice and sub-layer moments for the atomistic and synthetic cases respectively. \textbf{b}, Schematic of multilayer stacks together with measurement circuitry. \textbf{c}, (Top) Static magnetoresistance measurements as a function of magnetic field $B$ applied along $\phi =60~\text{deg}$. (Bottom) magnetoresistance for $\phi =60~\text{deg}$ and $B=55~\text{mT}$ after we excited the device with varied dc current ($I_\text{dc}$), displaying its $I_\text{dc}$ independence. \textbf{d}, Two-dimensional colour plot for rectified voltages $V$ with varying both microwave frequency $f$ and $B$ for $\phi =60~\text{deg}$. Dotted curves are calculated resonance peak positions for acoustic (dash) and optical (dash-dot) modes.  \textbf{e}, Individual $V$ measurements at 4 GHz after exciting the device with different $I_\text{dc}$ (purple from deep to light for -15 mA, 6 mA and 7 mA, yellow for 15 mA, respectively). \textbf{f}, The amplitude of anti-symmetric lineshape extracted by fitting Eq.~(\ref{fit}) to each $V$ trace for sweeping $I_\text{dc}$. \textbf{g}, Schematics of magnetisation dynamics of two coupled moments, $\bm{m_1}$ (represented in blue) and $\bm{m_2}$ (green), precessing in-phase for two time-independent Néel vector $\langle \bm{n}\rangle$. The Kittel vectors ($\bm{m}=(\bm{m_1} + \bm{m_2})/2$) for two cases are shown by the yellow and purple arrows. \textbf{h}, Time-dependent changes of an ac current and resistances for two $\langle \bm{n}\rangle$ cases. \textbf{i}, Resultant time-varying rectification voltages ($V(t)$) with $V_\text{dc}$ as its time-averaged component. Note that the rectified voltages are zero for the high-symmetry axes, i.e. 0, 90, 180 and 270 degrees.}
    \label{fig:Nswitching}
\end{figure}
The electrical detection of magnetic parameters is a fundamental cornerstone of modern spintronics, as it enables the transduction of spin-dependent information into measurable electrical signals that underpin a wide range of spin-based computing technologies\cite{BHATTI_MatToday2017,Dieny_NElec2020,Kurebayashi_arXiv2025}. The detection efficiency typically scales with the spatially averaged magnetisation, making it inherently more challenging to sense magnetic states in antiferromagnets than in ferromagnets. Over the past decade, however, remarkable progress has been made in the electrical detection of Néel vectors in antiferromagnets (defined as $\bm{n}=(\bm{m_1} - \bm{m_2})/2$, where $\bm{m_1}$ and $\bm{m_2}$ are the sublattice magnetisations), addressing a central challenge in the realisation of practical spintronic devices based on antiferromagnetic materials\cite{Jungwirth_NNano2016,Baltz_RMP2018}.

Anisotropic magnetoresistance (AMR) that is characterised by an even function of the magnetisation exists in antiferromagnets\cite{Shick_PRB2010} at both electron tunnelling\cite{Park_NMat2011} and bulk transport regimes\cite{Marti_NMat2014,Fina_NComm2014}. The planar Hall effect (PHE) which shares the same origin with AMR is another means for detecting current-induced Néel vector switching in CuMnAs\cite{Wadley_Science2016} and Mn$_2$Au\cite{Bodnar_NComm2018} where the spin-orbit interaction allows effective fields that are oppositely polarised (staggered) to act on sub-lattice moments individually. Low-frequency second-harmonic components based on the AMR/PHE mechanisms have been implemented for measuring Néel vectors in collinear antiferromagnets\cite{Godinho_NComm2018}. Néel vectors of insulating antiferromagnets can be probed by spin-Hall magnetoresistance (SMR) in an adjacent heavy metal layer\cite{Fischer_PRB2017,Hoogeboom_APL2017,Baldrati_PRB2018}. At the ultrafast timescale, optical pump-probe techniques have been routinely employed to study the time-domain magneto-optical characteristics of Néel vector dynamics\cite{Kirilyuk_RMP2010,Kimel_Nature2004,Saidl_NPhoton2017}.

Magnetic resonance is another widely used method for characterising magnetic properties\cite{Farle_RepProgPhys1998,stancil_book2009}. In this technique, the angular dependence of the resonance field reflects the magnetic anisotropy, while the linewidth provides insight into magnetic relaxation processes. In conventional ferromagnetic resonance (FMR), the amplitude of resonance peaks, proportional to the imaginary part of the magnetic susceptibility, has received less attention since vectorial information of magnetic dynamics is normally lost in amplitude.
In electrically-driven phase-sensitive spin-torque FMR (ST-FMR)\cite{Tulapurkar_Nature2005,Liu_PRL2011}, however,
resonance peak amplitudes yield valuable information about the excitation mechanisms, including the type of torques (field-like and damping-like), its magnitude and orientation\cite{Fang_NNano2011,Kurebayashi_NNano2014,MacNeill_NPhys2017}, offering opportunities for understanding the microscopic origins.
Recent applications of magnetic resonance techniques to antiferromagnets have produced exciting results, such as sub-terahertz spin pumping from insulating antiferromagnets\cite{Vaidya_Science2020}, the observation of nonlinear magnon dynamics driven by terahertz Néel spin-orbit torques\cite{Behovits_NComm2023}, and control of spin current polarisation via spin-wave interference\cite{Sheng_NPhys2025}. These advances in electrically reading and controlling antiferromagnetic states are promising for high-frequency applications, but further development of magnetic resonance techniques capable of direct and sensitive detection of antiferromagnetic order is critical. Addressing this would unlock new opportunities for integrating antiferromagnets into functional spintronic devices.

In this work, we demonstrate the electrical detection of the Néel vector by magnetic resonance.
In lithographically patterned micro-bars from synthetic antiferromagnets with Ta/NiFe/Ru/NiFe/Ta layer stacks, electric currents generate effective fields that exert magnetic torques on each NiFe layer in opposite directions, which we call synthetic staggered fields.
These fields acquire the staggered nature from the symmetric layer sequence, yet act analogously to atomistic staggered fields reported in CuMnAs\cite{Wadley_Science2016} and Mn$_2$Au\cite{Bodnar_NComm2018} due to the spin-orbit interaction in specific crystalline symmetries (Fig.~\ref{fig:Nswitching}a).
While the Néel vector in our devices cannot be easily probed by longitudinal magnetoresistance measurements, its time-varying components under magnetic resonance are highly sensitive to it, detectable via rectification techniques.
From the angular dependence of the rectified voltages, we determine not only 180$^{\circ}$ Néel vector switching but also the Néel vector direction.
Our micromagnetic analysis identifies the presence of defect-induced magnetic states appearing after current-induced Néel vector switching in our samples.
The efficiency of Néel vector switching is quantified in terms of the required current density; we \rev{obtain 7.4 $\times$ 10$^9$ A/m$^2$ with assumption of 1 nm top oxidation layer, which is} highly efficient compared with other reported values. We also demonstrate that the current density scales positively with the size of the in-plane magnetic anisotropy strength that is varied by device geometry.
The phase-sensitive scheme is applicable to a wide variety of antiferromagnets\cite{Jungwirth_NNano2016,Baltz_RMP2018} that exhibit magnetoresistance (MR) either intrinsically in conductive materials or via spin-Hall magnetoresistance for insulating antiferromagnets.

\section*{Néel vector switching measured by magnetic resonance}

We grew Ta/NiFe/Ru/NiFe/Ta multi-layers and lithographically patterned the films into micro-bar devices with which various transport measurements have been performed (Fig.~\ref{fig:Nswitching}b). We first show longitudinal MR measurements with sweeping the external magnetic field ($B$) in Fig.~\ref{fig:Nswitching}c - the change in longitudinal resistance ($R_\text{xx}$) can be explained by a model incorporating different types of MR mechanisms as explained in Supplementary Note 1. $R_\text{xx}$ at a fixed field strength of 55 mT after exciting the device with different dc currents ($I_\text{dc}$) exhibits constant resistance as shown in Fig.~\ref{fig:Nswitching}c, seemingly indicating that $I_\text{dc}$ is unable to perturb the magnetic states measured via dc MR measurements. On the same device, we switched the current excitation from dc to microwave, and electrically measured magnetic resonance by the ST-FMR scheme\cite{Tulapurkar_Nature2005,Fang_NNano2011,Liu_PRL2011}. A time($t$)-varying microwave current $I_{\rm mw}\sin\left( 2\pi ft \right) $ at frequency $f$ in the devices exerts spin torques on the coupled magnetic moments in NiFe layers, leading to magnetisation precession that is largely enhanced at resonance. This results in $t$-dependent resistance oscillation due to MR and causes the current rectification to produce $t$-averaged and $t$-dependent components ($V_\text{dc}$ and $V_\text{ac}$), from which we measure the former in this study. Two main (acoustic and optical) modes are recognised in a two-dimensional colour plot of $V_\text{dc}$ as a function of $B$ and $f$ in Fig.~\ref{fig:Nswitching}d, consistent with previous work using similar stacks\cite{Sud_PRB2020,Sud_APL2021,Sud_PRL2025}. We then applied $I_\text{dc}$ to the device without $B$ and removed it before starting each FMR scan, and discovered that the amplitude of $V_\text{dc}$ is $I_\text{dc}$-dependent, as we show some examples in Fig.~\ref{fig:Nswitching}e. For quantitative analysis, individual curves are analysed by the following Lorentzian functions to decompose their symmetric ($V_\text{sym}$) and anti-symmetric ($V_\text{asy}$) components:
\begin{equation}
        V_\text{dc} = V_\text{sym}\frac{\Delta B^2}{(B - B_\text{res})^2 + \Delta B^2} + V_\text{asy}\frac{\Delta B(B -  B_\text{res})}{(B - B_\text{res})^2 + \Delta B^2}.
        \label{fit}
    \end{equation}
Here, $B_\text{res}$ and $\Delta B$ are resonance field and half width at half maximum linewidth of resonance respectively. We used this equation for extracting the parameters (see Supplementary Figure 2 for fit examples), and the plot of $V_\text{asy}$ in Fig.~\ref{fig:Nswitching}f displays hysteresis, suggesting that $I_\text{dc}$ changes magnetic states in a non-volatile manner, which cannot be captured in static measurements (Fig.~\ref{fig:Nswitching}c) for the same excitation protocol. This is intriguing since although both measurements are performed via the identical two-probe contacts, they yield distinct responses, allowing for different magnetic parameters to be accessed.

We now explain the appearance of hysteresis using a phenomenological macrospin model with the Landau-Lifshitz-Gilbert (LLG) equation for exchange-coupled moments~\cite{Sud_APL2021}. Solving it (see Supplementary Note 2 for more details) leads to the calculated amplitude of time-dependent resistance oscillation for the acoustic mode $R_{\rm asy}^{\rm ac}$ given by:
\begin{equation}
    R_{\rm asy}^{\rm ac}(\langle n\rangle,\phi) =   \frac{\Delta R_{\rm MR} B_{\rm FL}}{8  \Delta B_{\rm ac} }  \frac{\langle n\rangle (1-2\langle n\rangle^2)}{\sqrt{1-\langle n\rangle^2}}\sin 2\phi \sin \phi .
    \label{Rasy}
\end{equation}

\noindent where $\Delta R_{\rm MR}$, $B_{\rm FL}$, $\Delta B_{\rm ac}$ and $\phi$ are the magnitude of MR change, the strength of the current-induced field-like effective fields, the FMR linewidth of the acoustic mode in terms of magnetic field and the angle between the applied magnetic field and the current flow direction (see Fig.~\ref{fig:Nswitching}a), respectively; $\langle n\rangle$ is the time-averaged Néel vector component along the direction perpendicular to $B$ within the film plane - see Fig.~\ref{fig:Nswitching}g.
$R_{\rm asy}^{\rm ac}$ is odd under the reversal of $\langle n\rangle$, i.e. $R_{\rm asy}^{\rm ac}(\langle n\rangle,\phi)=-R_{\rm asy}^{\rm ac}(-\langle n\rangle,\phi)$, as schematically represented in Fig.~\ref{fig:Nswitching}h. $R_{\rm asy}^{\rm ac}\sin\left( 2\pi ft \right)$ is multiplied by the $\langle n\rangle$-independent $I_{\rm mw}\sin\left( 2\pi ft \right) $ to appear in our $V_\text{dc}$ measurements (Fig.~\ref{fig:Nswitching}i) with the characteristic sign change due to the $\langle n\rangle$ reversal. This mechanism is consistent with our observation in Fig.~\ref{fig:Nswitching}e-f, and the reproducibility of this sign reversal is further shown in the symmetric voltage component (Supplementary Figure 3), the optical resonance mode (Supplementary Figure 4) and using different devices (Supplementary Figure 5) as well as scanning nitrogen-vacancy magnetometry (Supplementary Note 9).

\begin{figure}[b!]
    \centering
    \includegraphics[width=0.98\linewidth]{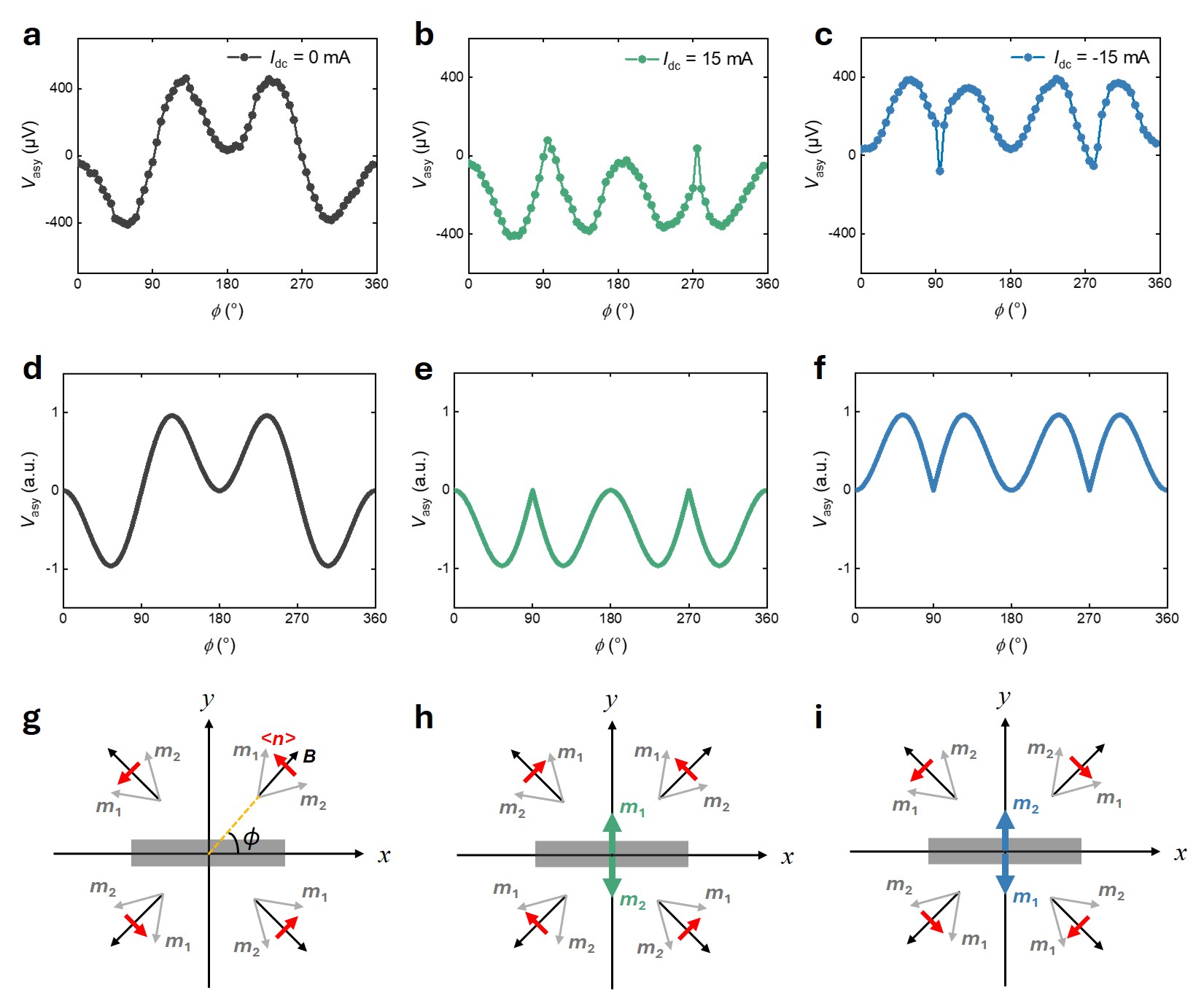}
    \caption{\textbf{Néel vector direction measured by magnetic resonance.} \textbf{a-c}, Experimentally measured $V_\text{asy}$ as a function of $\phi$ for $I_{\mathrm{dc}} = 0~\text{mA}$ (\textbf{a}), $I_{\mathrm{dc}} = 15~\text{mA}$ (\textbf{b}) and $I_{\mathrm{dc}} = -15~\text{mA}$ (\textbf{c}). \textbf{d-f}, Angular dependence of normalised $V_\text{asy}$ calculated for a constant $\langle n\rangle$ (\textbf{d}), $\langle n\rangle$ sign reversal applied for $\phi = 90^\circ$ - $270^\circ$ (\textbf{e}), and $\langle n\rangle$ sign reversal for $\phi = 0^\circ$ - 90$^\circ$ and 270$^\circ$ - 360$^\circ$ (\textbf{f}). \textbf{g-i}, Schematics of angular dependence of Néel vector states for the three cases: 0 mA (\textbf{g}), -15 mA (\textbf{h}) and 15 mA (\textbf{i}).}
    \label{fig:angular}
\end{figure}

\section*{Sensing Néel vector directions and switching mechanisms}
The present method can electrically determine the Néel vector orientation from the magnetic resonance. As proof of concept the angular dependence of $V_{\mathrm{asy}}$ was measured for three different $I_{\mathrm{dc}}$ (0, 15 and -15 mA) as shown in Fig.~\ref{fig:angular}a-c.

For $I_{\mathrm{dc}}$ = 0 mA, $V_{\mathrm{asy}}$ follows $\sin2\phi\sin\phi$, consistent with our model's prediction for a constant $\langle n\rangle$ in Eq.~\ref{Rasy}, plotted in Fig.~\ref{fig:angular}d.
Upon $I_{\mathrm{dc}}$ application, both angular dependences in Figs.~\ref{fig:angular}b-c show characteristic changes that are explained as follows.
Based on $\sin2\phi\sin\phi$, we realise that the angular dependence for $I_{\mathrm{dc}}$ = 15 mA can be reproduced by a sign-flip of the field angle window between 90$^\circ$ and 270$^\circ$ (Fig.~\ref{fig:angular}e). This corresponds to the sign reversal of $\langle n\rangle$ by the positive current excitation for that region. Likewise, the sign flip has been observed for 0$^\circ$- 90$^\circ$ and 270$^\circ$- 360$^\circ$ for $I_{\mathrm{dc}}$ = -15 mA, indicating the reversal of Néel vector for these windows (Fig.~\ref{fig:angular}f).

\begin{figure}[b!]
    \centering
    \includegraphics[width=1\linewidth]{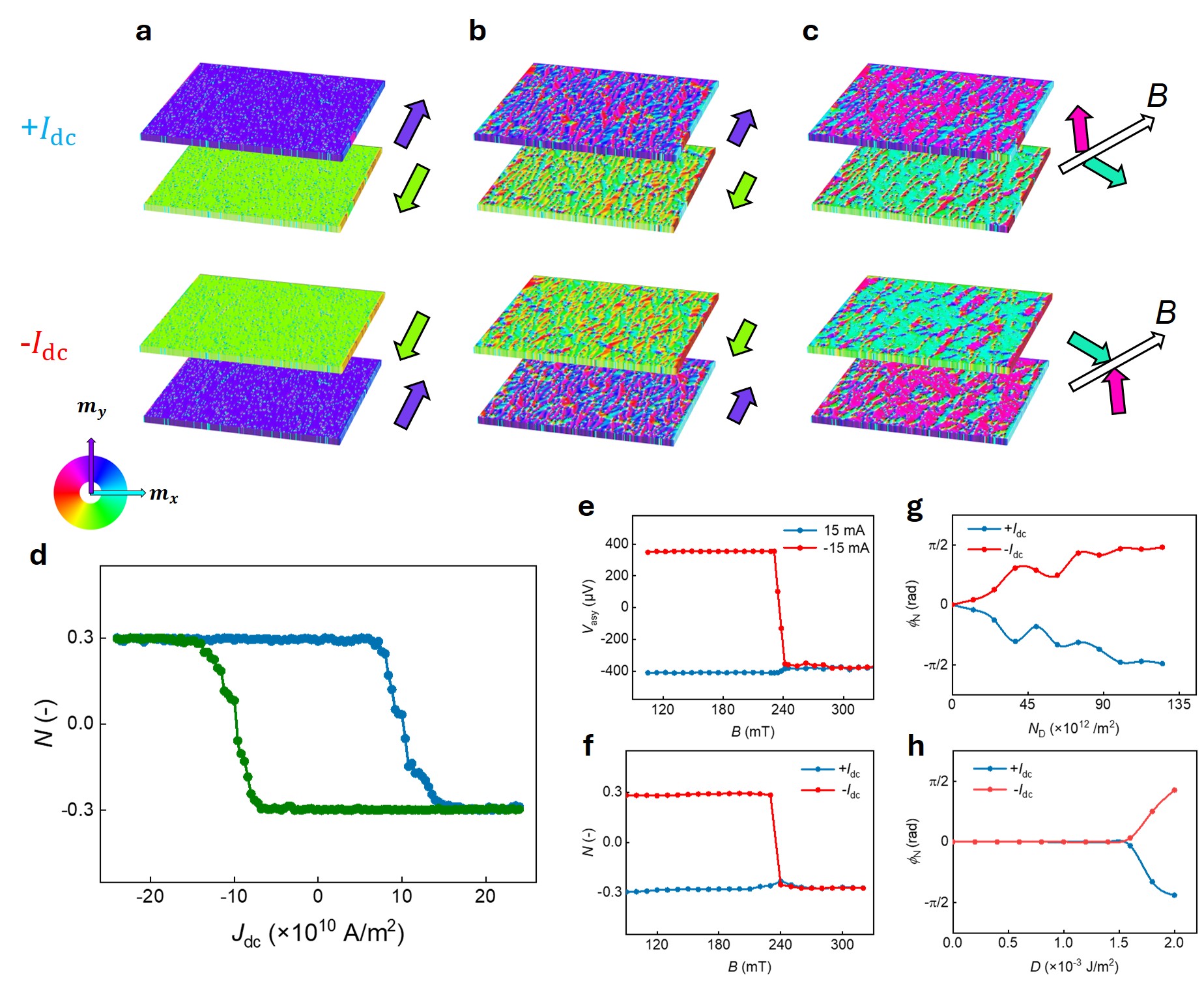}
    \caption{\textbf{Multi-domain magnetic textures in exchange-coupled layer systems.} \textbf{a-c}, Simulated magnetisation components maintained during current excitation for two different current polarities (\textbf{a}), after excitation (\textbf{b}), and after application of the external magnetic field (\textbf{c}). \textbf{d}, Calculated Néel components proportional to rectified voltages in our experiments, as a function of current density. \textbf{e}, Experimentally measured $V_\text{asy}$ as a function of magnetic field to represent the erasure of the meta-stable state memory. \textbf{f}, Calculated Néel components proportional to rectified voltages for the same condition as Fig.3\textbf{e}.  \textbf{g-h}, Néel vector direction calculated for different simulation parameters i.e. the number of defects $N_\text{D}$ (\textbf{g}) and DMI parameter $D$ (\textbf{h}). }
    \label{fig:efficiency}
\end{figure}

Since $V_{\mathrm{asy}}$ in ST-FMR measurements represents the spatially-averaged magnetic state in our device, we first apply the macrospin picture to explain the angular dependence using schematics in Figs.~\ref{fig:angular}g-i. For the zero current case, $\langle n\rangle$ is maintained constant with respect to $B$ during the measurements (Fig.~\ref{fig:angular}g) to produce the $\sin2\phi\sin\phi$ dependence. Now, using the experimental results, we reverse $\langle n\rangle$ for $\phi$ between 90$^\circ$ and 270$^\circ$ to map the angular dependence of $\langle n\rangle$ for $I_{\mathrm{dc}}$ = 15 mA as shown in Fig.~\ref{fig:angular}h. As detailed in Supplementary Note 3, this angular dependence in the canted state can be realised when we assume that the application of $I_{\mathrm{dc}}$ = 15 mA creates the two macrospin moments polarised along the $y$ axis as represented by the green arrows - imagine that each canted state in the four quadrants can be generated when applying $B$ on the initial state. For the negative current case, $\langle n\rangle$ is reversed from the positive current counterpart for the whole angle region, leading to the reversal of the two macrospin moments in the initial state as we show by blue arrows in Fig.~\ref{fig:angular}i. Both initial states produced by $I_{\mathrm{dc}}$ are consistent with the picture of synthetic staggered magnetic fields that orient along the $y$ axis but polarised oppositely for each magnetic layer. When the strength of the synthetic staggered fields is sufficient (i.e. large $I_{\mathrm{dc}}$) to overcome the energy barrier defined by the in-plane magnetic shape anisotropy, both moments can be polarised along the $y$ axis while $I_{\mathrm{dc}}$ is on. However, this is inconsistent with the simple macrospin approximation, since the $y$ axis is magnetically hard within the plane due to its shape anisotropy, meaning that as soon as we turn off $I_{\mathrm{dc}}$ as we do in our experiments before starting FMR scans, the energetically high states cannot be sustained and relaxed to the low-energy states with the $x$ polarisation. This indicates that more complex processes are required to explain our experimental observation.

\begin{figure}[b!]
    \centering
    \includegraphics[width=0.6\linewidth]{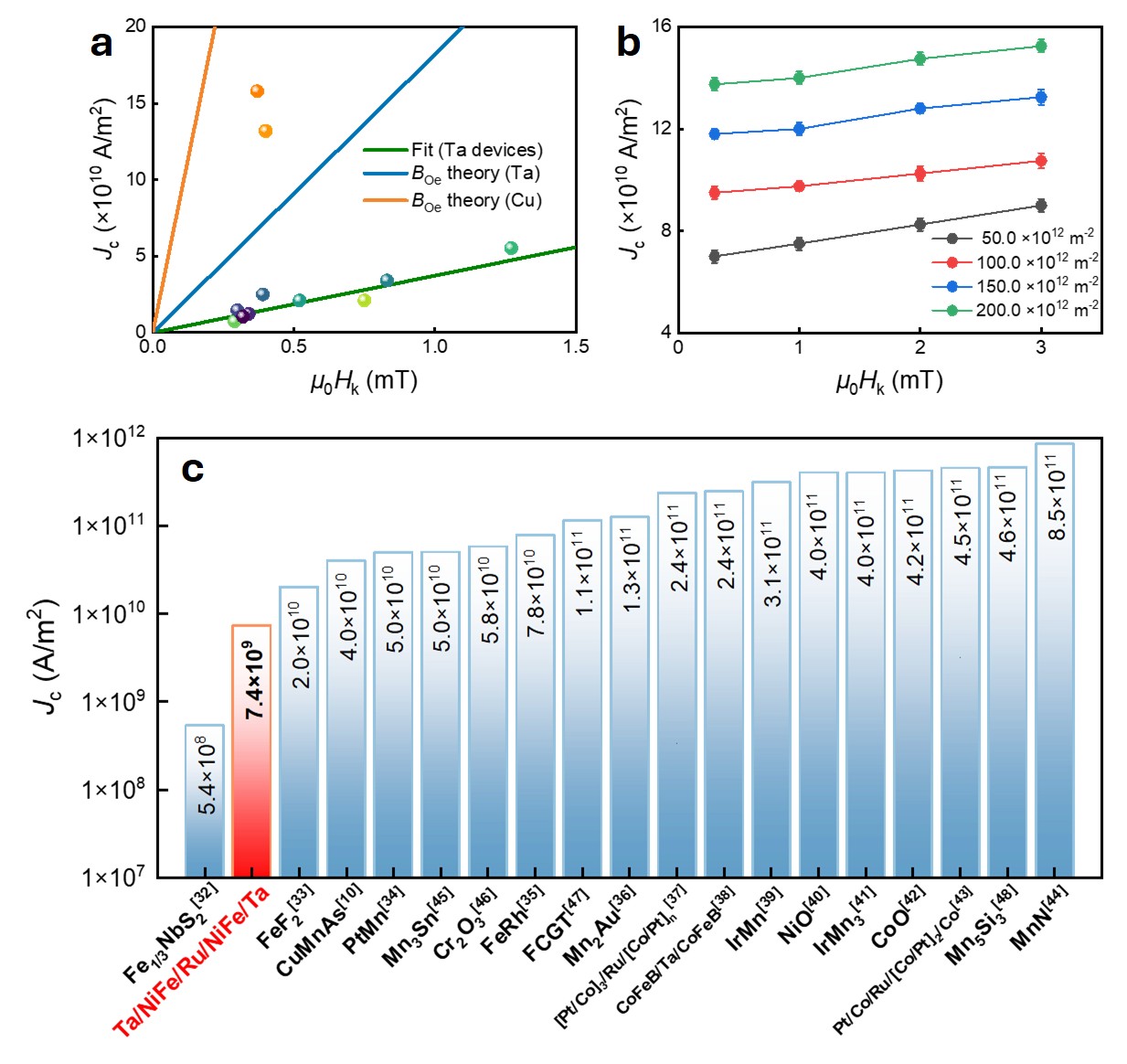}
    \caption{\textbf{Current-driven Néel vector switching efficiency.} \textbf{a}, Experimentally extracted critical current density for the Néel vector switching as a function of uniaxial anisotropy field. 9 Ta-based devices (blue to purple) and 2 Cu-based devices (orange) were measured. \rev{Solid lines each represent calculated Oersted fields ($B_\text{Oe}$) with thickness reduction due to possible oxidation of the 1 nm top (Cu or Ta) layer, for Ta-based (blue) and Cu-based devices (orange), respectively.} \rev{The green line corresponds to the average switching efficiency of the Ta-based devices ($2.68\times10^{-14}$~T$\cdot$m$^2$/A).} Detailed data for all devices can be found in Supplementary Note 4-6. \textbf{b}, Simulated critical current as a function of anisotropy field. \textbf{c}, Comparison of critical current densities for different material systems (field free \cite{Wadley_Science2016, nair2020electrical_32,ajejas2023current_33, duttagupta2020spin_34,wu2022current_37,bodnar2019imaging_39,zhu2020spin_40,shi2017spin_41,bai2025current_42,chen2018antidamping_43,arpaci2021observation_44,yang2022current_45,bi2017anomalous_46,dunz2020spin_48} and field assisted \cite{hu2022efficient_35,he2024electrical_36,guo2025layer_38,han2024electrical_47}, respectively). \rev{In \textbf{a} and \textbf{c}, $J_\text{c}$ is calculated assuming 1 nm oxidation of the top layer}.}
    \label{fig:5}
\end{figure}

We therefore consider a more realistic scenario of multi-domain states, incorporating the presence of magnetic defects that are randomly distributed across the entire sample area (see details in Supplementary Figure 6). Structural defects have been shown to act as domain-wall pinning sites that govern the micromagnetic states in antiferromagnets\cite{Wadley_NNano2018,bodnar2019imaging_39,Reimers_NComm2022}. Here we examine the role of artificial defects for interlayer exchange-coupled magnetic layers.
The mechanisms of current-induced Néel vector switching in our devices proceed via two stages: (1) Néel vector rotation driven by $I_{\mathrm{dc}}$, and (2) subsequent formation of fragmented micro-domain states after current is switched off.

In our micromagnetic simulations, we can reproduce the $y$-polarised states as shown in Fig.~\ref{fig:efficiency}a.
We found that the removal of $I_{\mathrm{dc}}$ leads to fragmented multi-domain states (Fig.~\ref{fig:efficiency}b) that emerge due to the combined effects of structural defects and interfacial Dzyaloshinskii–Moriya interaction (DMI).
Remarkably, a canted state generated by applying a subsequent magnetic field follows the pattern of the original multi-domain states as observed in Fig.~\ref{fig:efficiency}c, suggesting the non-volatile memory property of the current-induced magnetic textures. It is this property that results in the hysteresis observed experimentally (i.e. Fig.~\ref{fig:Nswitching}f), which is reproduced by the simulations shown in Fig.~\ref{fig:efficiency}d that plots the $\langle n\rangle$-dependent term in Eq.~(\ref{Rasy}), i.e. $N=\langle n\rangle (1-2\langle n\rangle^2)/\sqrt{1-\langle n\rangle^2}$. Furthermore, our micromagnetic simulations can reproduce the angular dependence of $V_\text{asy}$ in Fig.~\ref{fig:angular}, as shown in Supplementary Figure 7. The multi-domain states are robust and sustained until we apply the magnetic field required for the fully-polarised state in which the sub-moment exchange takes place at zero energy costs. This picture is supported both experimentally (Fig.~\ref{fig:efficiency}e) and numerically (Fig.~\ref{fig:efficiency}f) by observing the rectified voltages that collapse into one value for magnetic fields greater than 240 mT which corresponds to the magnitude of the interlayer exchange coupling strength. The mechanism is deterministic, providing a fully controllable method to either retain or erase the current-induced Néel vector states.

A series of systematic simulations varying key parameters identified the mechanisms generating fragmentation into multi-domain states. We simulated domain configurations following strong $\pm I_{\mathrm{dc}}$ excitation as a function of defect density and calculated the angle between the spatially-averaged Néel vector orientation and the longitudinal bar axis ($\phi_\text{N}$), which is plotted in Fig.~\ref{fig:efficiency}g. As expected, higher defect density correlates with larger $\vert \phi_\text{N}\vert$, with sign reversal of $\phi_\text{N}$ by the current polarity. We attribute this trend to defects acting as pinning sites that impede the formation of fully relaxed states locally. Figure~\ref{fig:efficiency}h reveals that interfacial DMI strength is the other important parameter to form the fragmented multi-domain states with the polarisation along the y direction.

\section*{Switching efficiency}

The Néel vector switching efficiency is characterised by the critical current density $J_\text{c}$ that is defined by $I_\text{dc}$ at which the Néel vector reversal is completed\rev{, divided by the conductive cross-section of the device. Since the top capping layer is partially oxidised, we calculate $J_\text{c}$ assuming that its top 1 nm is oxidised (Supplementary Notes 4 and 6)}. Since the Néel vector rotation and subsequent switching occur when the current-driven staggered effective fields overcome the anisotropy energy barrier, we expect a positive correlation between $J_\text{c}$ and the magnitude of the in-plane anisotropy strength. To examine this experimentally, we patterned micro-bar devices with various lateral aspect ratios to control the uniaxial anisotropy field ($\mu_0H_\text{k}$) that can be experimentally quantified by the in-plane angular dependence of $B_\text{res}$ (see more details in Supplementary Note 5).
 An experimentally extracted $J_\text{c}$-$\mu_0H_\text{k}$ plot from nine Ta-based devices (Fig.~\ref{fig:5}a) shows that Néel vector switching requires larger $J_\text{c}$ for more anisotropic devices, in agreement with the proposed Néel vector switching mechanism. Micromagnetic simulations further present a positive $J_\text{c}$-$\mu_0H_\text{k}$ correlation, with $J_\text{c}$ increasing with defect density due to the role of pinning. This positive scaling persists in simulations extended to much larger anisotropy fields, up to 20~mT, supporting the applicability of this switching mechanism to devices with stronger magnetic anisotropy (Supplementary Figure 8).

To the best of our knowledge, the $J_\text{c}$ of \rev{7.4} $\times$ 10$^{9}$ A/m$^2$ observed in our study (Supplementary Figure 9) is the smallest value reported for full reversal of the Néel vector at room temperature, with the only demonstration of a lower $J_\text{c}$ in Fe$_{1/3}$NbS$_2$ below 42 K\cite{nair2020electrical_32}, as summarised in Fig.~\ref{fig:5}c. We confirmed that this switching current density is independent of the applied pulse width from 1~s to 100~$\mu $s and that Joule heating does not significantly alter the device's magnetic anisotropy (Supplementary Note 7). It deterministically produces 180$^\circ$ Néel vector switching as evident from the sign reversal of FMR peaks in $V_\text{dc}$, which maintains its state over several days without any sign of relaxation (see Supplementary Figure 10). Therefore we consider it to be non-volatile against thermal fluctuation, which can only be destroyed by applied magnetic fields that overcome the interlayer exchange coupling as shown in Fig.~\ref{fig:efficiency}e. Such states can be accessible via electric currents that produce field-like staggered magnetic fields, analogous to the spin-orbit driven fields in crystalline antiferromagnets~\cite{Wadley_Science2016, Bodnar_NComm2018}, but here realised in synthetic antiferromagnets through stack-sequence engineering by design.
The importance of the field-like nature of the staggered magnetic fields is further supported by successful switching experiments in multilayer stacks of Cu/NiFe/Ru/NiFe/Cu where anti-damping torques are largely suppressed due to the weak spin-Hall effect in Cu (Supplementary Figure 11).

\section*{Conclusion}
We present a versatile methodology for electrical detection of Néel vector switching based on magnetic resonance techniques. The current-induced staggered effective fields acting differently on each magnetic layer induce the 180$^\circ$ reversal of Néel vectors, which results in a hysteresis in the rectified voltages due to a $\pi$ phase change in the time-dependent precessional motion. The angular dependence of the FMR peak amplitude calls for a model beyond the macrospin limit to explain the magnetisation polarised along the direction perpendicular to the magnetic easy axis. Our micromagnetic simulations with artificial defects as well as appropriate DMI parameters successfully address this. The Néel vector switching mechanism is the competition between the in-plane uniaxial anisotropy and the current-induced staggered fields, supported by experimental evidence that the critical current density for switching scales positively with the in-plane shape anisotropy field.
The approach enables nanoscale characterisation of antiferromagnetic order under electrically driven conditions, offering a new platform for exploring defect-mediated spin textures and their role in magnonic transport. Beyond fundamental insight, the demonstrated efficiency and scalability suggest strong prospects for integrating antiferromagnetic logic and memory functionalities into high-frequency on-chip spintronic architectures.

\section*{Methods}
\subsection*{Sample growth}
The SAF films used in this study were grown on thermally oxidised Si substrates using magnetron co-sputtering techniques at a base pressure of $3\times10^{-7}$ Pa. The SAF samples with stacking patterns of Ta(5)/NiFe(5)/Ru(0.2-0.4)/NiFe(5)/Ta(5) (in nm) and an extra Au contact layer on top were patterned into rectangular bars with varied length/width using the standard photolithography and Ar ion milling technique. Detailed profile of all samples can be found in Supplementary Note 4-6. The top Ta layer is expected to be partially oxidised after growth. However, we confirm that this oxidation does not significantly affect the magnetic properties of the NiFe layer underneath, since we observe synthetic antiferromagnetic behaviour, indicating that the two NiFe layers have similar magnetic properties. This is further confirmed by vibrating sample magnetometry (VSM), which shows a fully compensated magnetic moment at zero field (Supplementary Figure 1)

\subsection*{Measurement technique}
For ST-FMR measurements, the sample was placed on a coplanar waveguide, on which each side of a micro-bar device was wire-bonded to the waveguide signal line or ground line of the waveguide. The microwave current and dc current were applied to the signal line through a bias-tee. A lock-in technique was used to improve the signal-to-noise ratio: the microwave current and the lock-in amplifier were pulse modulated at a reference frequency (13 Hz). The difference in dc voltage across the sample was measured.

\subsection*{Micromagnetic simulations}
Micromagnetic simulations were performed using the Mumax3 software package, with the material parameters set as follows: saturation magnetisation, $6.2\times10^{\,5}$ A/m; exchange constant, $1.0\times10^{-11}$ J/m; damping factor, $0.1$; interfacial DMI, $1.9\times10^{\,-3}$ J/m$^{2}$ . The cell size was $5\times5\times5$ nm$^{3}$. The grid size was $800 \times 400 \times 5$. Simulation results can be found in Supplementary Figures 6 and 7.

\section*{Data availability}

The data presented in the main text and the Supplementary Information are available from the corresponding authors upon reasonable request.

\section*{Acknowledgements}
The authors thank Takeshi Seki and Del Atkinson for providing their comments to the manuscript. H.K. thanks the Leverhulme Trust for financial support via their research fellowship (RF-2024-317), and the Engineering and Physical Sciences Research Council (EPSRC) for support via grant EP/X015661/1. D.P. is supported by the EPSRC and SFI Centre for Doctoral Training in Advanced Characterisation of Materials Grant Ref: EP/S023259/1. A. S. thanks JSPS Postdoctoral fellowship for research in Japan (P21777).  S. M. thanks to CSRN in CSIS at Tohoku University and JSPS KAKENHI (Nos. 21H04648, 21H05000, 22F21777 and 22KF0030) and X-NICS of MEXT (No. JPJ011438). S. F. thanks to JSPS KAKENHI (Nos. 24H00039 and 24H02235) and X-NICS of MEXT (No. JPJ011438). \rev{C. B. and R. S. thank for financial support from the UK government Department for Business, Innovation, Science and Trade through: National Measurement System (NMS) funding (semiconductor metrology); and the UK national Quantum Technologies program (quantum enabled nanoscale magnetic field metrology).}

\section*{Author Contributions}
H.K. conceived the research project; A.S. and S.I. grew the multilayer thin-films and patterned them into micro-bar devices; A.S. initially discovered the Néel vector switching phenomena in ST-FMR, and Y.H., Z.C. and J.B. continued further magnetoresistance and ST-FMR measurements, and performed data analysis with support by H.K., C.Y., Z.X., C.W.F.F., D.P., T.W. and M.C.; Micromagnetic simulations were carried out by Y.H. with help by T.D.; C.B. and R.S. performed NV measurements; H.K., S.F. and S.M. supervised this project; H.K., Y.H. and Z.C. prepared the manuscript draft, and all authors contributed to its revision.

\clearpage
\raggedbottom
\makeatletter
\@fleqnfalse
\makeatother
\setcounter{figure}{0}
\setcounter{table}{0}
\setcounter{equation}{0}
\renewcommand{\thefigure}{S\arabic{figure}}
\renewcommand{\theHfigure}{S\arabic{figure}}
\renewcommand{\theequation}{S\arabic{equation}}
\renewcommand{\theHequation}{S\arabic{equation}}
\renewcommand{\theHtable}{S\arabic{table}}
\thispagestyle{empty}
{\raggedright\sffamily\bfseries\fontsize{20}{25}\selectfont Supplementary Information:\\
Synthetic staggered fields switch N\'eel vectors probed by magnetic resonance\par}
\vskip25pt
\begin{center} {\Large\textbf{CONTENTS}} \end{center}
{\large Supplementary Note 1: Magnetoresistance in Ta/NiFe/Ru/NiFe/Ta multi-layers}\vspace{0.2cm} \\
{\large Supplementary Note 2: Phenomenological model for voltage rectifications} \vspace{0.2cm} \\
{\large Supplementary Note 3: Macrospin simulations for Néel vectors as a function of dc currents}\vspace{0.2cm} \\
{\large Supplementary Note 4: Quantification of current distribution and current-induced torques and effective fields}\vspace{0.2cm} \\
{\large Supplementary Note 5: Data for quantifying the in-plane magnetic anisotropy for different devices}\vspace{0.2cm} \\
{\large Supplementary Note 6: Switching efficiency and driving force}\vspace{0.2cm} \\
{\large Supplementary Note 7: Pulse width and assessment of Joule-heating effects} \vspace{0.2cm} \\
{\large Supplementary Note 8: Scaling of Joule heating and thermal stability with reduced dimensions} \vspace{0.2cm} \\
{\large Supplementary Note 9: Scanning nitrogen-vacancy magnetometry}\vspace{0.2cm} \\
{\large Supplementary Figure 1: VSM data for synthetic antiferromagnets} \vspace{0.2cm} \\
{\large Supplementary Figure 2: Fitting of ST-FMR spectra}\vspace{0.2cm} \\
{\large Supplementary Figure 3: The current-induced Néel vector reversal probed by symmetric component for acoustic mode}\vspace{0.2cm} \\
{\large Supplementary Figure 4: The current-induced Néel vector reversal probed by optical mode}\vspace{0.2cm} \\
{\large Supplementary Figure 5: The current-induced Néel vector reversal measured in other devices}\vspace{0.2cm} \\
{\large Supplementary Figure 6: Micromagnetic simulation of defect-controlled domain structures}\vspace{0.2cm} \\
{\large Supplementary Figure 7: Micromagnetic simulation of angular dependence}\vspace{0.2cm} \\
{\large Supplementary Figure 8: Scaling of critical switching current density with higher shape anisotropy field}\vspace{0.2cm} \\
{\large Supplementary Figure 9: Switching efficiency in device No.2}\vspace{0.2cm} \\
{\large Supplementary Figure 10: Non-volatility of current-induced magnetic states}\vspace{0.2cm} \\
{\large Supplementary Figure 11: Control experiments with Cu/NiFe/Ru/NiFe/Cu multi-layers}\vspace{0.2cm} \\

\newpage

\section*{Supplementary Note 1: Magnetoresistance in Ta/NiFe/Ru/NiFe/Ta multi-layers}
\begin{figure}[H]
    \centering
    \includegraphics[width=0.9\linewidth]{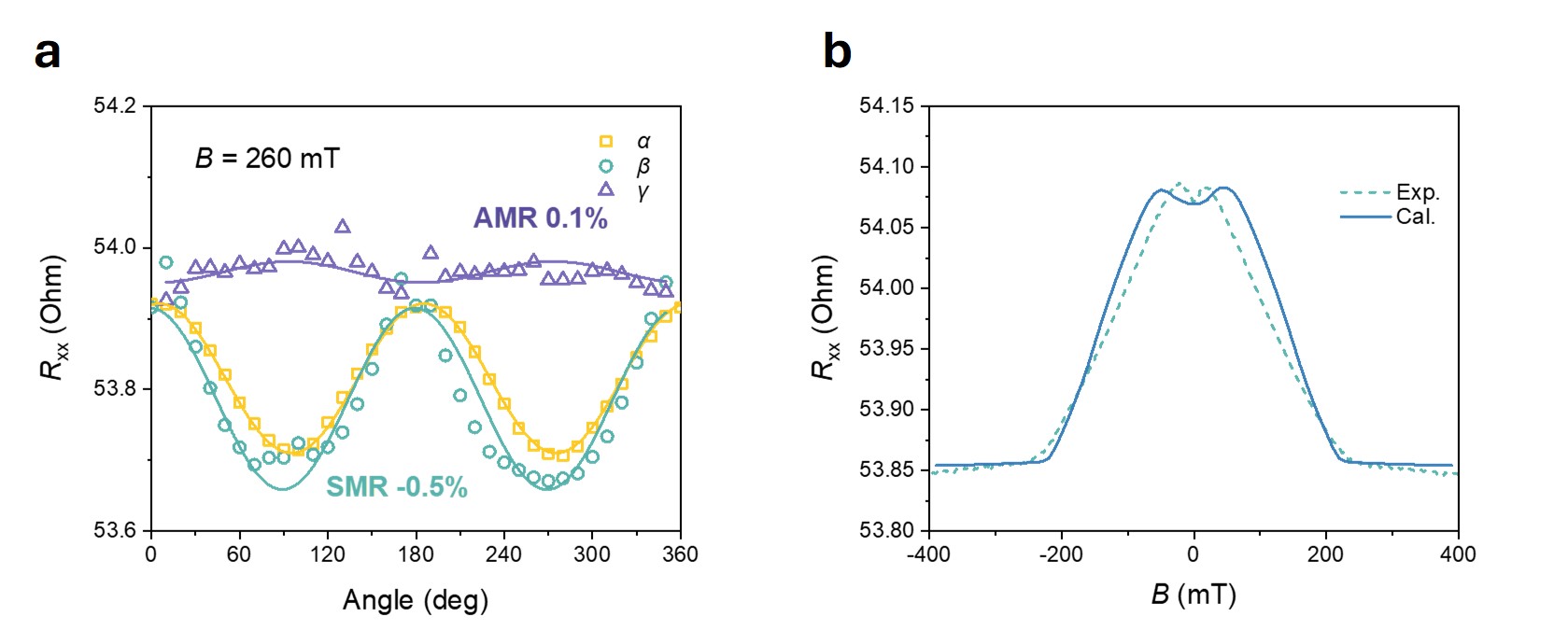}
    \caption{\textbf{a}, Angular dependence of magnetoresistance for the $x$-$y$ ($\alpha$), $y$-$z$ ($\beta$), $x$-$z$ ($\gamma$) plane rotation with 260 mT external field. \textbf{b}, Static magnetoresistance measurement and calculation as a function of magnetic field $B$ applied along $\phi =60~\text{deg}$.}
    \label{fig:MR}
\end{figure}

\begin{figure}[H]
    \centering
    \includegraphics[width=0.45\linewidth]{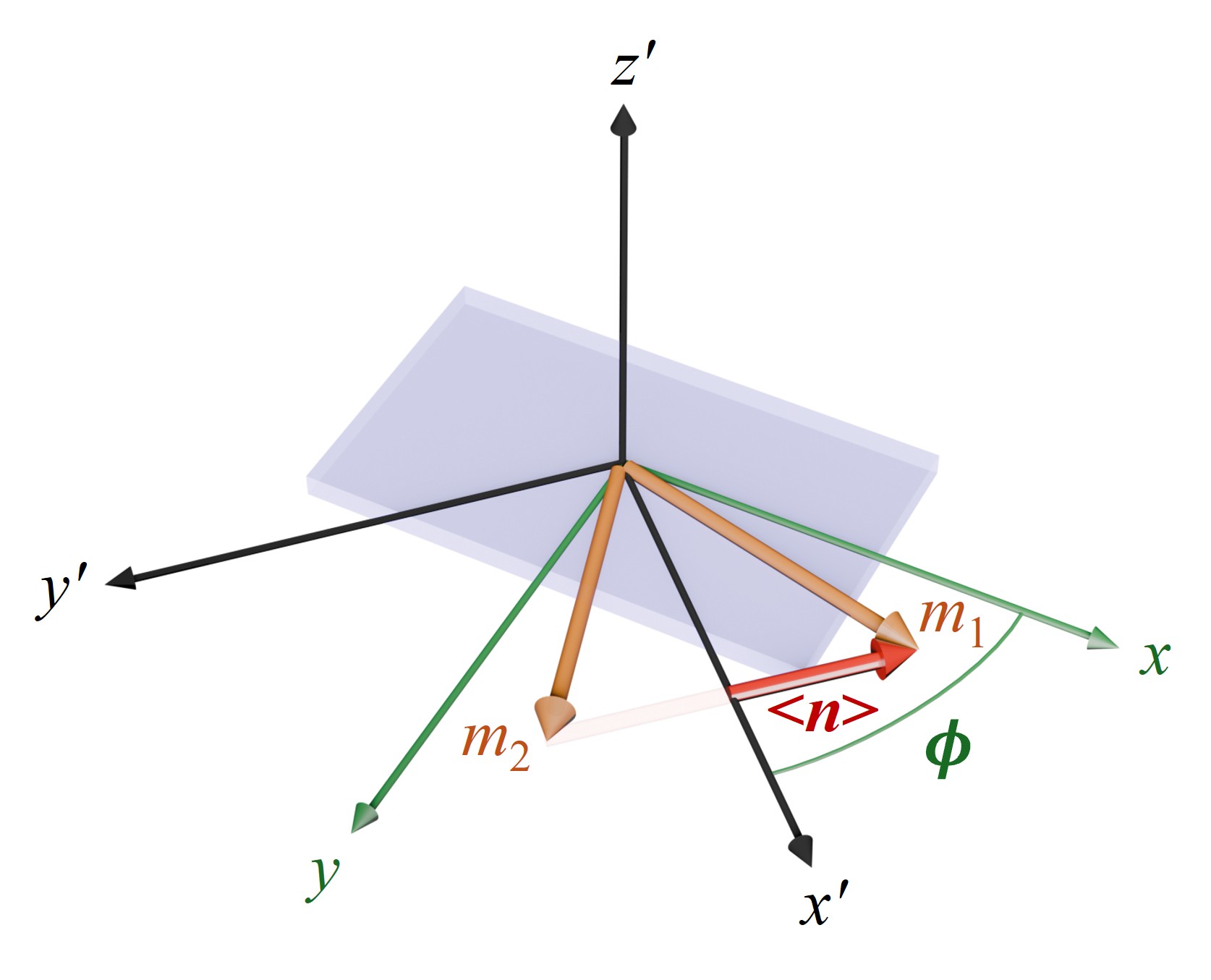}
    \caption{Coordination system used in our derivation. In this geometry, we have two Cartesian coordinate systems defined by the current flow ($x$) direction and the external magnetic field ($x'$) direction. The angle between the two is defined by $\phi$ and the Néel vector is defined by $\langle n\rangle$.}
    \label{fig:Coordination}
\end{figure}

 The anisotropic magnetoresistance (AMR) and spin Hall magnetoresistance (SMR) contributions can be extracted by the angular dependence of fixed-magnetic-field magnetoresistance (MR) in the $x$-$z$ and $y$-$z$ planes (Fig.~\ref{fig:MR}a), respectively, and the following phenomenological expression: $R_\text{total} = R_0 + \Delta R_\text{AMR} \cdot ((m_{1}\cdot \hat{e}_{x})^2 + (m_{2}\cdot \hat{e}_{x})^2)/2 +\Delta R_\text{SMR} \cdot ((m_{1}\cdot \hat{e}_{y})^2 + (m_{2}\cdot \hat{e}_{y})^2)/2 + \Delta R_\text{GMR} \cdot \int (1-m_i \cdot m_j)/2 \, dA$.
 Here, $R_0$, $\Delta R_\text{AMR}$, $\Delta R_\text{SMR}$ and $\Delta R_\text{GMR}$ are the resistance independent from the MR mechanisms and resistance changes for AMR, SMR and giant-magnetoresistance, respectively; $\hat{e}_i$ is the unit vector expression for each axis $i$; $m_i$ and $m_j$ are the normalised moments of each pair of spins in area $A$ between the first and second ferromagnetic layers attained from the Mumax3 multidomain simulation as mentioned in the main-text. The $x$-$y$-$z$ coordinates are defined in Fig.~\ref{fig:Coordination}.
The oscillation amplitude of each out-of-plane angular dependence leads to $\Delta R_\text{AMR} = 0.001 \cdot R_{0}=0.054~\text{Ohm}$ and $\Delta R_\text{SMR} = -0.005 \cdot R_{0}= -0.27~\text{Ohm}$. $\Delta R_\text{GMR}= -0.08~\text{Ohm}$ is the giant magnetoresistance (GMR) component which we found from fitting. Note that these are effective parameters of the individual mechanisms. Figure~\ref{fig:MR}b displays in-plane field-sweep MR measurements ($\phi =60~\text{deg}$), together with a curve calculated by the MR model, indicating fair agreement between the two.
\newpage

\section*{Supplementary Note 2: Phenomenological model for voltage rectifications}

The rectification voltages arising from magnetisation dynamics in synthetic antiferromagnets are derived below.
For the two sublattice magnetisations $m_1=m^0_1+\delta m_1(t)$ and $m_2=m^0_2+\delta m_2(t)$, we define the Kittel and Néel vectors $m = \langle m\rangle+\delta m(t)=(m_1+m_2)/2$, respectively, $n=\langle n\rangle+\delta n(t)=(m_1-m_2)/2$. The Landau-Lifshitz-Gilbert(LLG) equations using these coupled moments are:
\begin{equation}
\begin{aligned}
\frac{dm}{dt} &= -\Omega_\text{L} m\times u + \Omega_\text{B}\Big[m\times\Big(m\cdot \hat{e}_z\Big)\hat{e}_z + n\times\Big(n\cdot \hat{e}_z\Big)\hat{e}_z\Big] + \alpha\left(m\times \frac{dm}{dt} + n\times \frac{dn}{dt}\right),
\label{eq:LLG_m}
\end{aligned}
\end{equation}
\begin{equation}
\begin{aligned}
\frac{dn}{dt} &= -\Omega_\text{L} n\times u + \Omega_\text{B}\Big[n\times\Big(m\cdot \hat{e}_z\Big)\hat{e}_z + m\times\Big(n\cdot \hat{e}_z\Big)\hat{e}_z\Big] + 2\Omega_\text{ex}\Big(n\times m\Big)+\alpha\left(n\times \frac{dm}{dt} + m\times \frac{dn}{dt}\right).
\label{eq:LLG_n}
\end{aligned}
\end{equation}
Here, $\gamma$ denotes the gyromagnetic ratio, $\Omega_\text{L}=\gamma B$ is the Larmor precession frequency associated with the externally applied magnetic field ($B$), $\Omega_\text{B}=\gamma B_\text{s}$ is the demagnetising frequency associated with the demagnetisation field ($B_\text{s}$), $\Omega_\text{ex}=\gamma B_\text{ex}$ is the exchange frequency corresponding to the interlayer exchange field ($B_\text{ex}$), and $u$ is the unit vector along the direction of $B$ which in this case is applied along $u=\hat{e}_{x'}$.
\\
Two configurations of a positive or negative Néel vector are considered: $m^0_1=(\sqrt{1-\langle n\rangle^2},\; \langle n\rangle,\; 0),\; m^0_2=(\sqrt{1-\langle n\rangle^2},\; -\langle n\rangle,\; 0)$ with linearised $m$ and $n$ as $m=(
\sqrt{1-\langle n\rangle^2} + \delta m_{x'}(t),\; \delta m_{y'}(t),\;  \delta m_{z'}(t))$ and \rev{$n=(\delta n_{x'}(t),\; \langle n\rangle + \delta n_{y'}(t),\; \delta n_{z'}(t))$}. Solving Eqs.~\eqref{eq:LLG_m} and \eqref{eq:LLG_n} for the first order term, we find the matrix form of two equations by using the Fourier transformation:
\rev{\begin{align}
\begin{pmatrix}
 i\Omega & (i\alpha\Omega + \Omega_\text{B} )\sqrt{1-\langle n\rangle^2} + \Omega_\text{L} & 0 \\
 -i\alpha\Omega \sqrt{1-\langle n\rangle^2} - \Omega_\text{L} & i\Omega & i\alpha\Omega \langle n\rangle \\
 0 & -(i\alpha\Omega +2\Omega_\text{ex}+\Omega_\text{B})\langle n\rangle\ & i\Omega
\end{pmatrix}
\begin{pmatrix}
\delta m_{y'}(t) \\ \delta m_{z'}(t) \\ \delta n_{x'}(t)
\end{pmatrix}
= \bm{\tau}_m,
\label{eq:matrix_m}
\end{align}}

\rev{\begin{align}
\begin{aligned}
\begin{pmatrix}
 i\Omega & 0 & -(i\alpha\Omega + \Omega_\text{B}) \langle n\rangle \\
 0 & i\Omega & (i\alpha\Omega - 2\Omega_\text{ex}+\Omega_\text{B})\sqrt{1-\langle n\rangle^2} + \Omega_\text{L} \\
 (i\alpha\Omega + 2\Omega_\text{ex})\langle n\rangle & (-i\alpha\Omega+2\Omega_\text{ex}) \sqrt{1-\langle n\rangle^2} - \Omega_\text{L}  & i\Omega
\end{pmatrix}
\\[6pt]
\begin{pmatrix}
\delta m_{x'}(t) \\ \delta n_{y'}(t) \\ \delta n_{z'}(t)
\end{pmatrix}
= \bm{\tau}_n.
\end{aligned}
\label{eq:matrix_n}
\end{align}}
The right hand side of the equations are magnetic torques on each coupled moment. The complex resonance frequencies for acoustic mode can be obtained by using the determinant of the $3\times3$ matrix on the left hand side, the relation $\sqrt{1-\langle n\rangle^2}=\Omega_\text{L}/2\Omega_\text{ex}$ is used to simplify the expression,
\begin{equation}
\tilde{\Omega}_\text{ac} = \sqrt{\Omega_\text{L}^2\left(1+\frac{\Omega_\text{B}}{2\Omega_\text{ex}}\right)} + i\frac{\alpha}{2}\left(\frac{\Omega_\text{L}^2}{2\Omega_\text{ex}}+\Omega_\text{B}+2\Omega_\text{ex}\right),
\end{equation}
In this model, an electrical current $I_0$ flows along the $x$ direction, $B$ is applied along $x'$ from which we define $\phi$ with respect to the $x$ direction. In this coordinate, \rev{a field-like effective field $B_\text{FL}$} is generated along the $y$ direction, exerting field-like torques ($\tau_\text{FL,1(2)}$) on $m_1$ and $m_2$ as
\begin{align}
\tau_\text{FL,1(2)} &= \gamma B_{\text{FL},1(2)} m_{1(2)} \times \hat{e}_{y} \nonumber\\
&= \gamma B_{\text{FL},1(2)} (\sqrt{1-\langle n\rangle^2}\cos\phi-\langle n\rangle\sin\phi)\, \hat{e}_{z'}.
\end{align}
For symmetric stacking, $B_{\text{FL},1}=-B_{\text{FL},2}$. Therefore, the field-like torques acting on $m$ and $n$ in Eqs.~~\eqref{eq:matrix_m} –~\eqref{eq:matrix_n} are
\rev{\begin{align}
\tau_m &= \gamma
\begin{pmatrix}
0 \\
- B_\text{FL}\,\langle n\rangle \sin\phi \\
0
\end{pmatrix},
\label{eq:tau_m_FL}
\end{align}
\begin{align}
\tau_n &= \gamma
\begin{pmatrix}
0 \\
0 \\
B_\text{FL}\,\sqrt{1-\langle n\rangle^2} \cos\phi
\label{eq:tau_n_FL}
\end{pmatrix}.
\end{align}}
In addition to these torques, damping-like torques are generated by the spin Hall effect (SHE). For current direction $x$, the spin polarization is $\sigma=\hat{e}_{y}$ and giving torques $\tau_\text{DL,i}= \gamma B_\text{DL,i} \big[m_i \times (\hat{e}_{y} \times m_i)\big], (i=1,2)$.
For symmetric stacks $B_\text{DL,1} = -B_\text{DL,2}$. Therefore, the damping-like torques acting on $m$ and $n$ are
\rev{\begin{align}
\tau^\text{DL}_m &= \gamma
\begin{pmatrix}
- B_\text{DL}\,\langle n\rangle \sqrt{1-\langle n\rangle^2} \sin\phi \\
0 \\
B_\text{DL}\,\langle n\rangle^2 \sin\phi
\end{pmatrix},
\label{eq:tau_m_DL}
\end{align}
\begin{align}
\tau^\text{DL}_n &= \gamma
\begin{pmatrix}
B_\text{DL}\,\langle n\rangle \sqrt{1-\langle n\rangle^2} \cos\phi \\
B_\text{DL}\,(1-\langle n\rangle^2) \cos\phi \\
0
\label{eq:tau_n_DL}
\end{pmatrix}.
\end{align}}
Using Eqs.~\eqref{eq:matrix_m}, ~\eqref{eq:matrix_n}, ~\eqref{eq:tau_m_FL}, ~\eqref{eq:tau_n_FL}, ~\eqref{eq:tau_m_DL}, and ~\eqref{eq:tau_n_DL}, the dynamic components of acoustic mode $\delta m_{y'}$ and $\delta n_{x'}$ become
\rev{\begin{align}
\delta m_{y'} &=
\frac{\gamma\big[\Omega^2 B_\text{DL}\,\langle n\rangle\sqrt{1-\langle n\rangle^2}\sin\phi + i\Omega(\Omega_\text{L}+\Omega_\text{B}\sqrt{1-\langle n\rangle^2})(B_\text{FL}\langle n\rangle\sin\phi)\big]}{-i\Omega(-\Omega_\text{ac}+i\delta_\text{ac}-\Omega)(\Omega_\text{ac}+i\delta_\text{ac}-\Omega)}, \\
\delta n_{x'} &=
\frac{\gamma\big[-\Omega^2 B_\text{DL}\langle n\rangle^2\sin\phi - i\Omega B_\text{FL}(\Omega_\text{B}+2\Omega_\text{ex})\langle n\rangle^2\sin\phi\big]}{-i\Omega(-\Omega_\text{ac}+i\delta_\text{ac}-\Omega)(\Omega_\text{ac}+i\delta_\text{ac}-\Omega)},
\end{align}}
Here $\tilde{\Omega}_\text{ac} = \pm \Omega_\text{ac} + i\delta_\text{ac}$. By calculating the residue at two poles ($- \Omega_\text{ac} + i\delta_\text{ac}$) and ($\Omega_\text{ac} + i\delta_\text{ac}$) using the residue theorem, $\text{Res}(f,c) = \lim_{z\to c}(z-c)f(z)$, we obtain simplified forms:
\begin{align}
\delta m_{y'}(\Omega) &= \frac{\gamma}{\Omega_\text{ac}+i\delta_\text{ac}-\Omega}(a_{m_y'}+i b_{m_y'})
+ \frac{\gamma}{-\Omega_\text{ac}+i\delta_\text{ac}-\Omega}(-a_{my'}+i b_{my'}), \\
\delta n_{x'}(\Omega) &= \frac{\gamma}{\Omega_\text{ac}+i\delta_\text{ac}-\Omega}(a_{n_x'}+i b_{n_x'})
+ \frac{\gamma}{-\Omega_\text{ac}+i\delta_\text{ac}-\Omega}(-a_{nx'}+i b_{nx'}),
\end{align}
with the following coefficients:
\rev{\begin{align}
a_{m_y'} &= \tfrac{1}{2} B_\text{FL}\,\frac{\Omega_\text{ac}}{\Omega_\text{L}}\langle n\rangle\sin\phi, \\
b_{m_y'} &= -\tfrac{1}{2} B_\text{DL}\langle n\rangle\sqrt{1-\langle n\rangle^2}\sin\phi, \\
a_{n_x'} &= -B_\text{FL}\frac{(\Omega_\text{B}+2\Omega_\text{ex})}{2\Omega_\text{ac}}\langle n\rangle^2\sin\phi, \\
b_{n_x'} &= \tfrac{1}{2} B_\text{DL}\,\frac{\Omega_\text{L}(\Omega_\text{B}+2\Omega_\text{ex})}{\Omega_\text{ac}^2}\langle n\rangle^2\sqrt{1-\langle n\rangle^2}\sin\phi,
\end{align}}
For acoustic mode, GMR has no time-dependent component as $m_1$ and $m_2$ precess in-phase. The time-varying resistances due to AMR and SMR are calculated as
\begin{align}
\Delta R_\text{AMR}(t) &= \tfrac{1}{4}\Delta R_\text{AMR}\Big[(m_1\cdot \hat{e}_{x})^2+(m_2\cdot \hat{e}_{x})^2\Big] \nonumber\\
&= \tfrac{1}{2}\Delta R_\text{AMR}\Big[\cos^2\phi(m_{x'}^2+n_{x'}^2)+\sin^2\phi(m_{y'}^2+n_{y'}^2)-2\sin\phi\cos\phi(m_{x'}m_{y'}+n_{x'}n_{y'})\Big] \nonumber\\
&= \Delta R_\text{AMR}\Big[\cos^2\phi \sqrt{1-\langle n\rangle^2}\delta m_{x'}(t) + \sin^2\phi \langle n\rangle\delta n_{y'}(t) -\sin\phi\cos\phi\Big (\sqrt{1-\langle n\rangle^2}\,\delta m_{y'}(t)+\langle n\rangle\delta n_{x'}(t)\Big)\Big]\\
\Delta R_\text{SMR}(t) &= \tfrac{1}{4}\Delta R_\text{SMR}\Big[(m_1\cdot \hat{e}_{y})^2+(m_2\cdot \hat{e}_{y})^2\Big] \nonumber\\
&= \tfrac{1}{2}\Delta R_\text{SMR}\Big[\sin^2\phi(m_{x'}^2+n_{x'}^2)+\cos^2\phi(m_{y'}^2+n_{y'}^2)-2\sin\phi\cos\phi(m_{x'}m_{y'}+n_{x'}n_{y'})\Big] \nonumber\\
&= \Delta R_\text{SMR}\Big[\sin^2\phi \sqrt{1-\langle n\rangle^2}\,\delta m_{x'}(t) + \cos^2\phi \langle n\rangle\delta n_{y'}(t) +\sin\phi\cos\phi\Big( \sqrt{1-\langle n\rangle^2}\,\delta m_{y'}(t)+\langle n\rangle\delta n_{x'}(t)\Big) \Big]
\end{align}
These allow to find the analytical expressions of rectification voltages for the acoustic mode ($V^\text{ac}_\text{asy}$ and $V^\text{ac}_\text{sym}$) by using Ohm's law and taking dc components which are:
\rev{\begin{align}
V^\text{ac}_\text{asy} &= -\frac{ \gamma}{2\delta_\text{ac}}(\Delta R_\text{AMR}-\Delta R_\text{SMR})\,I^2_0\frac{d}{dI}(a_{m_y'}\sqrt{1-\langle n\rangle^2}\sin\phi\cos\phi+a_{n_x'}\langle n\rangle\sin\phi\cos\phi)\\
&= -\frac{ I_0 B_\text{FL}}{8\Delta B_\text{ac}}(\Delta R_\text{AMR}-\Delta R_\text{SMR})\,\frac{\langle n\rangle}{\sqrt{1-\langle n\rangle^2} }(1-2\langle n\rangle^2) \sin 2\phi \sin\phi,
\label{eq:Vasy_ac}
\end{align}
\begin{align}
V^\text{ac}_\text{sym} &= -\frac{ \gamma}{2\delta_\text{ac}}(\Delta R_\text{AMR}-\Delta R_\text{SMR})\,I^2_0\frac{d}{dI}(b_{m_y'}\sqrt{1-\langle n\rangle^2}\sin\phi\cos\phi+b_{n_x'}\langle n\rangle\sin\phi\cos\phi)\\
&= \frac{I_0 B_\text{DL}}{8\Delta B_\text{ac} \sqrt{1+B_\text S/2B_\text{ex}}}\, (\Delta R_\text{AMR}-\Delta R_\text{SMR})\,\langle n\rangle (1-2\langle n\rangle^2) \sin 2\phi \sin\phi,
\label{eq:Vsym_ac}
\end{align}}
\newpage

\section*{Supplementary Note 3: Macrospin simulations for Néel vectors as a function of dc currents}
\begin{figure}[H]
    \centering
    \includegraphics[width=0.9\linewidth]{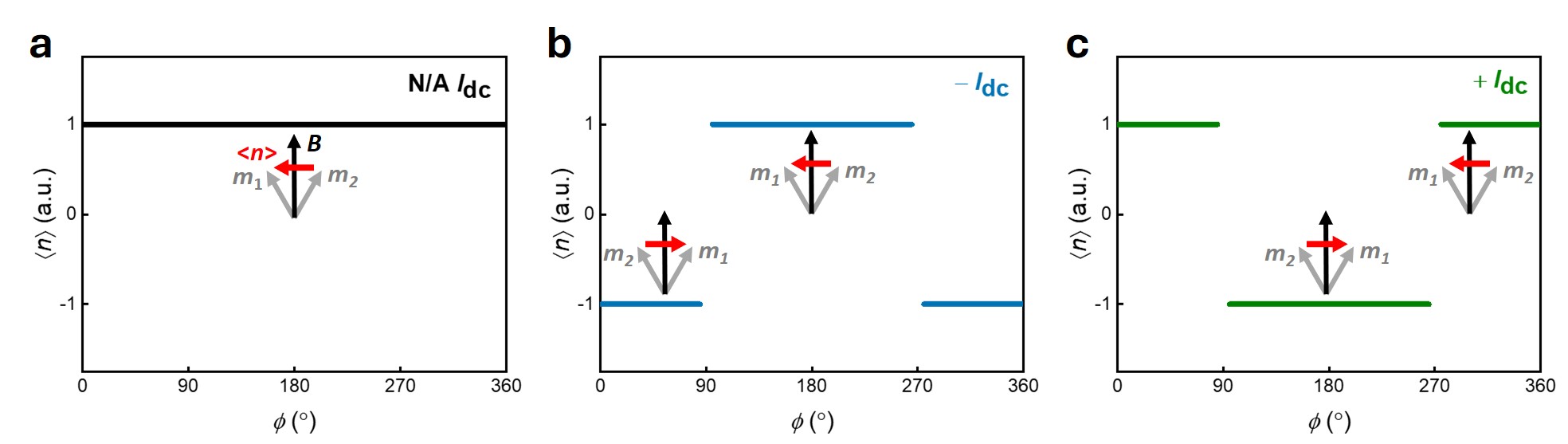}
    \caption{Macrospin simulation of Néel vector depending on $\phi$ for (\textbf{a}) initial state (no $I_\text{dc}$) and current induced state for $+I_\text{dc}$ (\textbf{b}) or $-I_\text{dc}$ (\textbf{c}). The macrospin simulation is carried out in Mumax3. }
    \label{fig:Macrospin_Neel_vector}
\end{figure}

Here we show our macrospin calculations for $\langle n\rangle$ when we apply $B$ for various directions within the film plane. In particular, we are interested in finding $\langle n\rangle$ when we initialise the two moments $m_1$ and $m_2$ by $I_\text{dc}$ = 0, -15 mA and 15 mA. For the last two cases, the current amplitudes are large enough such that the two moments are forced to be polarised along the $y$ direction and the current polarity reverses each moment by 180$^\circ$, i.e. the reversal of $\langle n\rangle$. We simulated the final state of $m_1$ and $m_2$ by applying $B$ = 50 mT for each angle and the equilibrium canted state was simulated for every 5$^\circ$ for each $I_\text{dc}$ case. As we can see, the $\langle n\rangle$ reversal as a function of $\phi$ is observed when we initialised with finite $I_\text{dc}$. These macrospin simulations results support the underlying picture that we illustrate in Figs.~2g-i.
\newpage

\section*{{Supplementary Note 4: Quantification of current distribution and current-induced torques and effective fields}}
In this section, we calculate the current distribution and estimate the current-induced effective fields in the Ta-based and Cu-based devices.
For a multilayer stack consisting of parallel conducting layers, the current distribution is determined by the conductance of each layer according to:
\begin{align}
 I_1:I_2:I_3:\cdots= G_1:G_2:G_3:\cdots=\sigma_1A_1:\sigma_2A_2:\sigma_3A_3:\cdots
\label{parallel model}
\end{align}
where $I_i$, $G_i$, $\sigma_i$, and $A_i$ denote the current, conductance, conductivity, and cross-sectional area of the $i$th layer within the stack, respectively. In this case, as all layers share the same width, Eq.~(\ref{parallel model}) simplifies to
\begin{align}
 I_1:I_2:I_3:\cdots= \sigma_1d_1:\sigma_2d_2:\sigma_3d_3:\cdots
\label{parallel model simp}
\end{align}
where $d_i$ is the thickness of layer $i$.
The conductivities of $\beta$-Ta, Ru, permalloy (Py) and Cu are taken as $1.0\times 10^{6}$ S/m, $1.4\times 10^{7}$ S/m, $2.8\times 10^{6}$ S/m and $2\times 10^{7}$ S/m, respectively. For the stack Ta[Cu](5nm)/ NiFe(5nm)/ Ru(0.4nm)/ NiFe(5nm)/ Ta[Cu](5nm), the calculated current distribution is
\begin{align}
I_\text{Ta}:I_\text{Py}:I_\text{Ru}:I_\text{Py}:I_\text{Ta}&=0.12:0.32:0.12:0.32:0.12
\label{distribution_Ta}
\end{align}
\begin{align}
I_\text{Cu}:I_\text{Py}:I_\text{Ru}:I_\text{Py}:I_\text{Cu}&=0.43:0.06:0.02:0.06:0.43
\label{distribution_Cu}
\end{align}
\rev{with the corresponding current density distribution as:
\begin{align}
J_\text{Ta}:J_\text{Py}:J_\text{Ru}:J_\text{Py}:J_\text{Ta}&=0.05:0.13:0.64:0.13:0.05
\label{density_distribution_Ta}
\end{align}
\begin{align}
J_\text{Cu}:J_\text{Py}:J_\text{Ru}:J_\text{Py}:J_\text{Cu}&=0.33:0.05:0.24:0.05:0.33
\label{density_distribution_Cu}
\end{align}}
Thus, the current flowing through a single Ta layer ($I_\text{Ta}$) or Cu layer ($I_\text{Cu}$) corresponds to approximately 12\% and 43\% of the total current ($I_\text{tot}$), respectively. \rev{The current densities in the single Ta layer ($J_\text{Ta}$) or Cu layer ($J_\text{Cu}$) are therefore $J_\text{Ta}=0.46 \ J_\text{tot}$ and $J_\text{Cu}=1.71 \ J_\text{tot}$, where $J_\text{tot}$ is the nominal current density calculated using the total thickness of the multilayer stack.}
\\
Here we also take into account the possibility of Ta oxidation within the top Ta layer. At room temperature, oxidation of Ta is self-limiting due to the formation of a dense passivating oxide layer. Consequently, only a very thin surface oxide on the order of a nanometer is expected on the 5 nm Ta capping layer~\cite{Stejskal_PRB2020}. \rev{Here, we consider 1 or 2 nm of Ta has been fully oxidised into insulating TaO$_{x}$, and the stack of the device is TaO$_{x}$(1nm)/ Ta(4nm)/ NiFe(5nm)/ Ru(0.4nm)/ NiFe(5nm)/ Ta(5nm) and TaO$_{x}$(2nm)/ Ta(3nm)/ NiFe(5nm)/ Ru(0.4nm)/ NiFe(5nm)/ Ta(5nm), respectively.} The current distribution within the system changes to:
\begin{align}
I_{\text{TaO$_{x}$}}:I_\text{Ta(4nm)}:I_\text{Py}:I_\text{Ru}:I_\text{Py}:I_\text{Ta(5nm)}&=0.00:0.09:0.33:0.13:0.33:0.12
\label{current_distribution_Ta_1nm}
\end{align}
\rev{\begin{align}
I_{\text{TaO$_{x}$}}:I_\text{Ta(3nm)}:I_\text{Py}:I_\text{Ru}:I_\text{Py}:I_\text{Ta(5nm)}&=0.00:0.07:0.34:0.13:0.34:0.12
\label{current_distribution_Ta_2nm}
\end{align}}
\rev{with the corresponding current density distribution as:
\begin{align}
J_\text{Ta(4nm)}:J_\text{Py}:J_\text{Ru}:J_\text{Py}:J_\text{Ta(5nm)}&=0.05:0.13:0.64:0.13:0.05
\label{Density_distribution_Ta_1nm}
\end{align}
\begin{align}
J_\text{Ta(3nm)}:J_\text{Py}:J_\text{Ru}:J_\text{Py}:J_\text{Ta(5nm)}&=0.05:0.13:0.64:0.13:0.05
\label{Density_distribution_Ta_2nm}
\end{align}}
\rev{Due to this oxidisation, the conductive thickness of the stack decreases from 20~nm to 19 (18)~nm, increasing the corresponding total current density ($J_\text{tot}$) for a given current by 5.3\% (11.1\%). Meanwhile, since both Ta layers have the same conductivity, they carry the same current density no matter of the oxidisation depth. The current density distribution within the stack shows negligible difference between each oxidisation thickness, thus its effect on the spin-Hall-effect-related quantities below is negligible. Throughout this work, we consider a more realistic 1~nm oxidisation of the top capping layer, $J_\text{tot}$ and the critical switching current density $J_\text{c}$ are therefore calculated using the 19~nm conductive thickness, in which case $J_\text{Ta(4nm)}= J_\text{Ta(5nm)}=0.45 \ J_\text{tot}$.}
\\
Next we consider the source of current-induced fields that may act on magnetic layers. Within this discussion, we estimate the strength of current-induced fields due to the spin-Hall effect as well as Oersted field, using a finite sheet model. For Ta-based devices, the spin Hall angle ($\theta_\text{SH}$) is quantified from Eq.~(\ref{eq:Vsym_ac}) using the amplitude of symmetric component ($V_\mathrm{sym}$) and linewidth ($\Delta B_\text{ac}$) as extracted from the ST-FMR spectra measured at $\phi=60^\circ$ shown in Fig.~(\ref{fig:Fitting_ac}):
\begin{align}
\theta_\text{SH}=\frac{2eM_sd_\mathrm{FM}B_\text{SHE}}{\hbar J_\text{Ta}}
=\frac{2eM_sd_\mathrm{FM}\Delta B_\text{ac}\sqrt{1+B_\text S/2B_\text{ex}}V_\mathrm{sym}}{\hbar J_\text{Ta}I_0(\Delta R_\text{AMR}-\Delta R_\text{SMR}) \langle n\rangle (1-2\langle n\rangle^2)\sin 2\phi \sin\phi}
=0.25
\label{theta_SH}
\end{align}
where $V_\mathrm{sym}=5.0\times10^{-5}$~V, $\Delta B_\text{ac}=4$~mT, $\hbar=1.05\times10^{-34}\mathrm{J\cdot s}$, $e=1.6\times10^{-19}$, $M_s=6.20\times10^5\mathrm{A/m}$, $I_0=28\ \mathrm{mA}$, \rev{$J_\text{Ta}=2.21\times10^{10} \mathrm{A/m^2}$}, $(\Delta R_\text{AMR}-\Delta R_\text{SMR})=0.32~\Omega$ and $\langle n\rangle=0.96$ at a $B_\text{res}=55\ \mathrm{mT}$.
Thus the damping-like effective field arising from the spin-Hall effect ($B_\text{SHE}$) is given by:
\begin{align}
 B_\text{SHE}=\frac{\hbar\theta_\text{SH}}{2ed_\text{Ta}M_s}J_\text{Ta}=2.64\times10^{-14}J_\text{Ta}=\rev{1.19}\times10^{-14}J_\text{tot} \ \mathrm{(T)}
\label{B_SHE}
\end{align}
The $B_\text{SHE}$ for Cu-based devices is not considered due to a negligible $\theta_\text{SH}$ in Cu. The effective field of field-like torques ($B_\text{FL}$) within Ta-based devices is quantified from Eq.~(\ref{eq:Vasy_ac}) using the amplitude of antisymmetric component ($V_\mathrm{asy}$) of ST-FMR spectra measured at $\phi=60^\circ$ shown in Fig.~(\ref{fig:Fitting_ac}):
\begin{align}
B_\text{FL}
&= \frac{ 8V^\text{ac}_\text{asy}\Delta B_\text{ac} }{I_0(\Delta R_\text{AMR}-\Delta R_\text{SMR})}\,\frac{\sqrt{1-\langle n\rangle^2}}{\langle n\rangle(1-2\langle n\rangle^2)\sin 2\phi \sin\phi}=1.81\times10^{-3}~\mathrm{(T)},
\label{B_FL}
\end{align}
where $V_\mathrm{asy}=3.2\times10^{-4}$~V, $\Delta B_\text{ac}=4$~mT, $\langle n\rangle=0.96$ and $I_0=28\ \mathrm{mA}$ which corresponds to \rev{$J_\text{tot}=4.91\times10^{10} \mathrm{A/m^2}$}. The $B_\text{FL}$ efficiency within this device is therefore given by: \rev{$B_\text{FL}/J_\text{tot}=3.68\times10^{-14}~\mathrm{T\cdot m^2/A}$}.\\
The current-induced Oersted fields ($B_\text{Oe}$) within both devices are estimated using the finite sheet model~\cite{Gauthier_2dOe}. \rev{Since our devices are much wider than the stack thickness, the Oersted field generated by the current $I_i$ in each conductive layer has a magnitude of $\mu_0 I_i/2W$, which is independent of distance and points in opposite directions above and below that layer.  The fields acting on the top and bottom NiFe layers are therefore:
\begin{align}
B_\text{Oe,1}&=\frac{\mu_0}{2W}\left(I_\text{Ru}+I_\text{Py,2}+I_\text{cap,2}-I_\text{cap,1}\right),\nonumber\\
B_\text{Oe,2}&=\frac{\mu_0}{2W}\left(I_\text{cap,2}-I_\text{cap,1}-I_\text{Py,1}-I_\text{Ru}\right),
\label{B_Oe_layers}
\end{align}
where $I_\text{cap,1}$ ($I_\text{Py,1}$) and $I_\text{cap,2}$ ($I_\text{Py,2}$) are the currents in the top and bottom capping (NiFe) layers, respectively, and $W$ is the device width. The N\'eel vector is rotated by the staggered component of these fields, which we define as
\begin{align}
B_\text{Oe}=\frac{B_\text{Oe,1}-B_\text{Oe,2}}{2}=\varsigma\frac{\mu_0I_\text{tot}}{2W}=\varsigma\frac{\mu_0d_\text{tot}}{2}J_\text{tot}, \quad \varsigma=\frac{I_\text{Py,1}+I_\text{Py,2}+2I_\text{Ru}}{2I_\text{tot}},
\label{B_Oe_finite_sheet}
\end{align}
where $\mu_0=4\pi\times10^{-7}\ \mathrm{T\cdot m/A}$ and $d_\text{tot}$ is the conductive stack thickness (19~nm).
The difference between the currents in the two outermost conductive layers (Cu or Ta), which arises from oxidation of the top one, produces the same field on both NiFe layers and cancels in Eq.~(\ref{B_Oe_finite_sheet}).
Such a common field has a negligible effect on the N\'eel vector rotation, as it is much smaller than the interlayer exchange field ($2B_\text{ex}=240$~mT in our case).
Oxidation therefore affects $\varsigma$ only through the redistribution of the current among the layers.
For Ta-based devices with 1~nm oxidation, Eq.~(\ref{current_distribution_Ta_1nm}) gives $\varsigma_\text{Ta}=0.46$.} For Cu devices where the top 1~nm is fully oxidised into insulating \rev{CuO$_{x}$}, the stack of device can be considered as \rev{CuO$_{x}$}(1nm)/ Cu(4nm)/ NiFe(5nm)/ Ru(0.4nm)/ NiFe(5nm)/ Cu(5nm), the current distribution within the system changes to:
\rev{\begin{align}
I_{\text{CuO}_x}:I_\text{Cu(4nm)}:I_\text{Py}:I_\text{Ru}:I_\text{Py}:I_\text{Cu(5nm)}&=0.00:0.37:0.07:0.02:0.07:0.47
\label{distribution_Cu_1nm}
\end{align}}
\rev{This gives $\varsigma_\text{Cu}=0.092$. $\varsigma$ changes only from 0.45 to 0.46 for Ta and from 0.084 to 0.092 for Cu}. Therefore, $B_\text{Oe}$ calculated from the finite sheet model are:
\begin{align}
 B_\text{Oe}(\text{Ta})/J_\text{tot}=\rev{5.49\times10^{-15}} \ \mathrm{T\cdot m^2/A}
\label{B_Oe_Ta}
\end{align}
\begin{align}
 B_\text{Oe}(\text{Cu})/J_\text{tot}=\rev{1.10\times10^{-15}} \ \mathrm{T\cdot m^2/A}
\label{B_Oe_Cu}
\end{align}

\newpage

\section*{Supplementary Note 5: Data for quantifying the in-plane magnetic anisotropy for different devices}
\begin{figure}[H]
    \centering
    \includegraphics[width=0.8\linewidth]{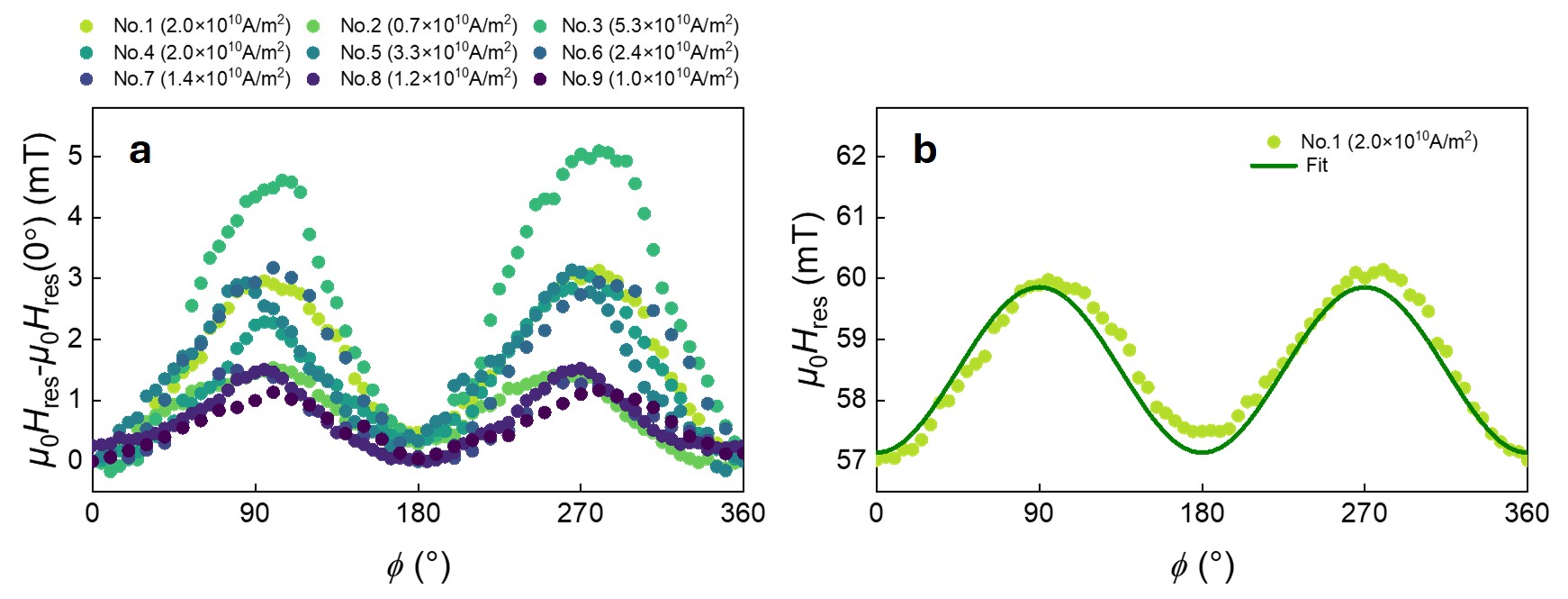}
    \caption{Angular dependence of resonant field ($B_\text{res}$) probed by ST-FMR for devices No. 1-9. All devices exhibit a characteristic two-fold symmetry with maxima at $\phi = 90^\circ$ and $270^\circ$ and minima at $\phi = 0^\circ$ and $180^\circ$, demonstrating that their magnetic energy is dominated by a uniaxial in-plane anisotropy aligned with the long axis of the SAF devices. The critical switching current of each devices are labelled, where the serial number of devices is consistent with Fig. 4 from the main text and supplementary figures.}
    \label{fig:anisotropy}
\end{figure}

To quantify the in-plane magnetic anisotropy of our SAF devices, we measured the angular dependence of resonance field ($B_\text{res}$) at a fixed microwave frequency. For an in-plane SAF with uniaxial (shape) anisotropy along bar direction, the coupled LLG equation can be written as:
\begin{equation}
\begin{aligned}
\frac{dm}{dt} &= -\Omega_\text{L} m\times u + \Omega_\text{B}\Big[m\times\Big(m\cdot \hat{e}_z\Big)\hat{e}_z + n\times\Big(n\cdot \hat{e}_z\Big)\hat{e}_z\Big]+ \Omega_\text{k}\Big[m\times\Big(m\cdot \cdot \hat{e}_k\Big)\hat{e}_k + n\times\Big(n\cdot \hat{e}_k\Big)\hat{e}_k\Big],
\end{aligned}
\end{equation}
\begin{equation}
\begin{aligned}
\frac{dn}{dt} &= -\Omega_\text{L} n\times u + \Omega_\text{B}\Big[n\times\Big(m\cdot \hat{e}_z\Big)\hat{e}_z + m\times\Big(n\cdot \hat{e}_z\Big)\hat{e}_z\Big]+ \Omega_\text{k}\Big[n\times\Big(m\cdot \hat{e}_k\Big)\hat{e}_k + m\times\Big(n\cdot \hat{e}_k\Big)\hat{e}_k\Big]+ 2\Omega_\text{ex}\Big(n\times m\Big).
\end{aligned}
\end{equation}
Here, $\Omega_\text{L}=\gamma B$, $\Omega_\text{B}=\gamma B_\text{s}$, $\Omega_\text{ex}=\gamma B_\text{ex}$. $\Omega_\text{k}=\gamma \mu_0H_\text{k}$ is the anisotropy frequency corresponding to the anisotropy field ($\mu_0H_\text{k}$). $u$ is the unit vector along the direction of $B$ which is applied along $u=\hat{e}_{x'}$. $\hat{e}_k$ is the direction of anisotropy field which is along $x$ and is defined by $\hat{e}_k=\cos\phi\ \hat{e}_{x'}-\sin\phi\ \hat{e}_{y'}$. Under small $\Omega_\text{k}$ approximation ($\Omega_\text{k}<<\Omega_\text{B}$) and drop the second order terms, the resonance condition for acoustic mode derived from the coupled LLG equation is:
\begin{equation}
\Big(\frac{\omega_\text{ac}}{\gamma}\Big)^2=
B_\text{res}^2\Big(1 + \frac{B_\text{s}}{2B_\text{ex}}\Big)+
{\mu_0H_\text{k}}\cos(2\phi)\Big(2B_\text{ex}+B_\text{s}\Big)\Big(1-\frac{B_\text{res}^2}{2B_\text{ex}^2}\Big).
\label{eq.Hrf_ac}
\end{equation}
where $B_\text{res}$ is the resonance field, $\mu_0H_\text{k}$ is the in-plane uniaxial (shape) anisotropy field,  $B_\text{s}=\mu_0M_\text{s}$ is the saturation magnetisation, and $B_\text{ex}=\mu_0H_\text{ex}$ is the interlayer exchange coupling field. $\mu_0H_\text{k}$ is therefore extracted from fitting of each angular dependence plotted in Fig.~\ref{fig:anisotropy} using Eq.~\eqref{eq.Hrf_ac}. The extracted $\mu_0H_\text{k}$ for different devices are listed below in Table~\ref{tab:all_device_profile}.
\newpage

\section*{{Supplementary Note 6: Switching efficiency and driving force}}
\begin{table}[H]
    \centering
    \includegraphics[width=1.0\linewidth]{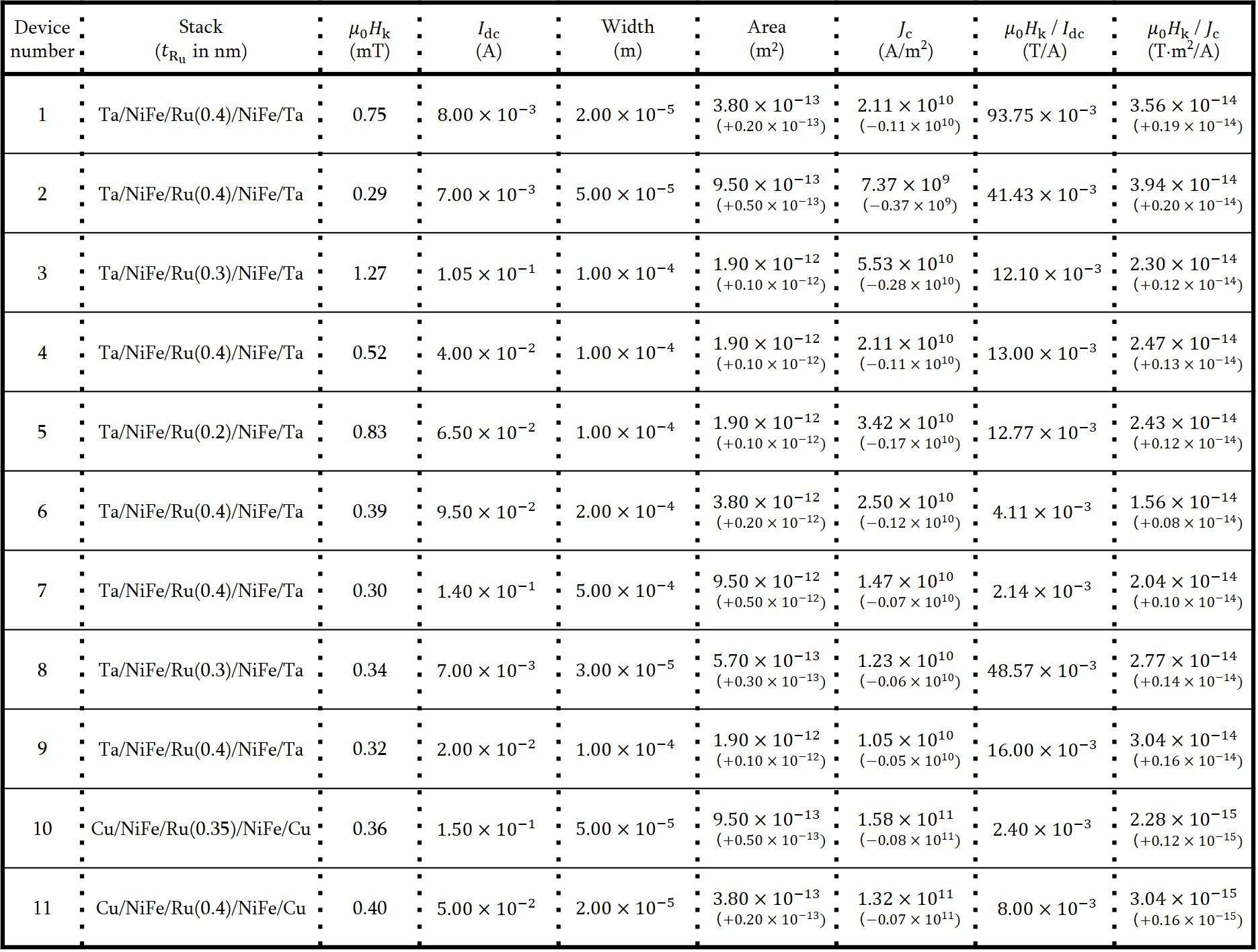}
    \caption{Table of all devices presented in this paper. Device No.\textbf{1-9} are Ta-based devices and No.\textbf{10, 11} are the Cu-based devices. The switching efficiency is defined by the magnitude of field generated per current ($\mu_0H_\text{k}/I_\text{dc}$) or per current density ($\mu_0H_\text{k}/J_\text{c}$). \rev{Data related to area and $J_c$ are calculated based on the assumption that the thickness of the whole stack is reduced by 1~nm due to potential oxidisation of the top capping layer, which are plotted in Fig. 4(a,c) in the main text. Numbers in bracket are calculated assuming the capping layer is not oxidised.} The serial number of devices is consistent with the main text and all other supplementary information.}
    \label{tab:all_device_profile}
\end{table}
\begin{figure}[H]
    \centering
    \includegraphics[width=1.0\linewidth]{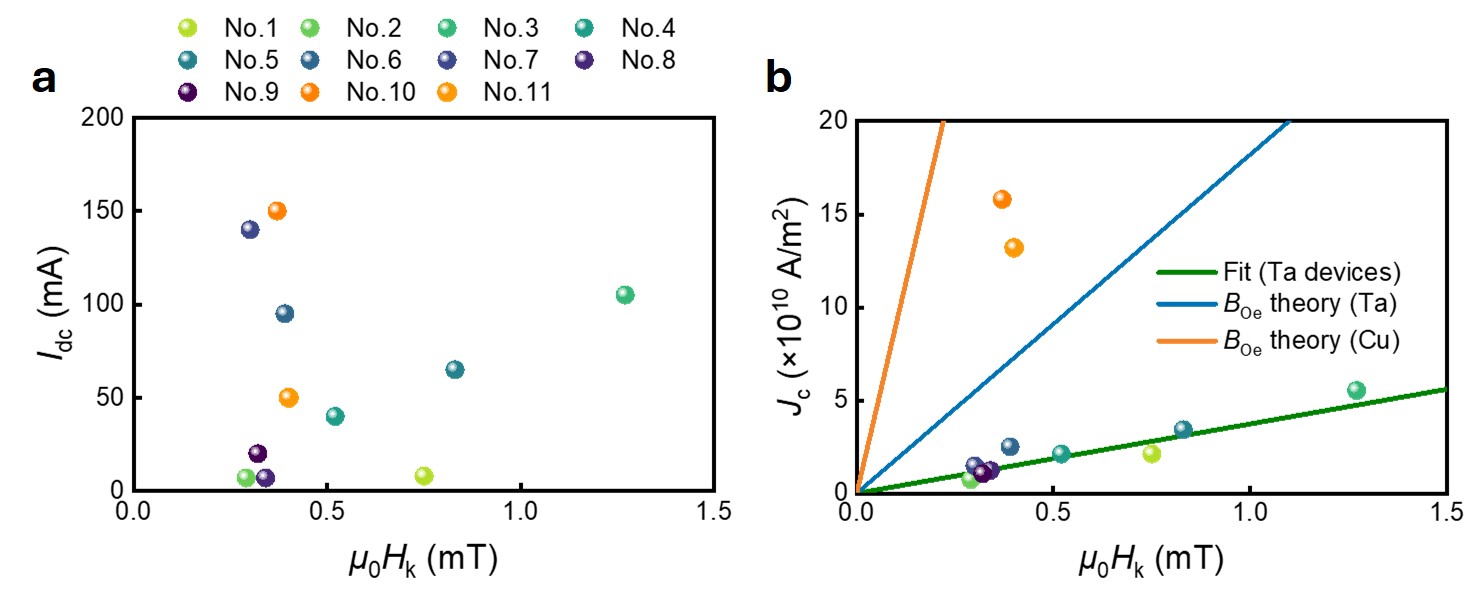}
    \caption{\textbf{a} Critical switching current ($I_\text{c}$) and \textbf{b} current density ($J_\text{c}$) as a function of shape anisotropy field ($\mu_0H_\text{k}$) for devices No.\textbf{1-11}. The switching efficiency predicted by Eqs.~\eqref{B_Oe_Ta} and ~\eqref{B_Oe_Cu} are compared with experimental results in \textbf{b}. \rev{$J_\text{c}$ in \textbf{b} is calculated assuming 1 nm oxidation of the top capping layer.} }
    \label{fig:anisotropy_scaling}
\end{figure}
To investigate the origin of current-induced switching in our SAF devices, nine Ta-based and two Cu-based devices are tested. The detailed device parameters are summarised in Table~\ref{tab:all_device_profile}. The dependence of critical switching current ($I_\text{c}$) and current density ($J_\text{c}$) on the shape anisotropy field ($\mu_0H_\text{k}$) is presented in Fig.~\ref{fig:anisotropy_scaling}. In contrast to a direct scaling with $I_\text{c}$, the Ta-based devices exhibit a clear correlation between $J_\text{c}$ and $\mu_0H_\text{k}$. However, as the Oersted field $B_\text{Oe}$ generated in the devices follows the finite sheet model (Eq.~\eqref{B_Oe_finite_sheet}), such scaling between $J_\text{c}$ and $\mu_0H_\text{k}$ is expected for switching driven by either $B_\text{Oe}$ or the field-like torque effective field $B_\text{FL}$.

The switching efficiency expected from a $B_\text{Oe}$ driven mechanism is calculated using Eqs.~\eqref{B_Oe_Ta} and \eqref{B_Oe_Cu}, and is compared with the experimental results in Fig.~\ref{fig:anisotropy_scaling}\textbf{b}. \rev{For Ta-based devices, the switching efficiency predicted by Eq.~\eqref{B_Oe_Ta} is \rev{$5.49\times10^{-15} \ \mathrm{T\cdot m^2/A} $}, which is approximately five times less efficient than the experimentally extracted average efficiency ($2.68\times10^{-14} \ \mathrm{T\cdot m^2/A} $)}. This substantial discrepancy suggests that the switching in Ta-based devices cannot be predominantly attributed to the Oersted field, but is instead more consistent with a mechanism driven by $B_\text{FL}$.

By contrast, the Cu-based devices (device No.10 and 11) exhibit a switching efficiency approximately one order of magnitude smaller than that of the Ta-based devices \rev{($2.28\times10^{-15}$ and $3.04\times10^{-15}~\mathrm{T\cdot m^2/A}$), which is within a factor of 2--3 of the $B_\text{Oe}$ driven prediction from Eq.~\eqref{B_Oe_Cu} ($1.10\times10^{-15}~\mathrm{T\cdot m^2/A}$)}. These results therefore indicate that current-induced switching in the Ta-based devices is primarily governed by \rev{$B_\text{FL}$}, whereas the switching in the Cu-based device is predominantly driven by $B_\text{Oe}$.

\newpage

\section*{Supplementary Note 7: Pulse width and assessment of Joule-heating effects}
\begin{figure}[H]
    \centering
    \includegraphics[width=1.0\linewidth]{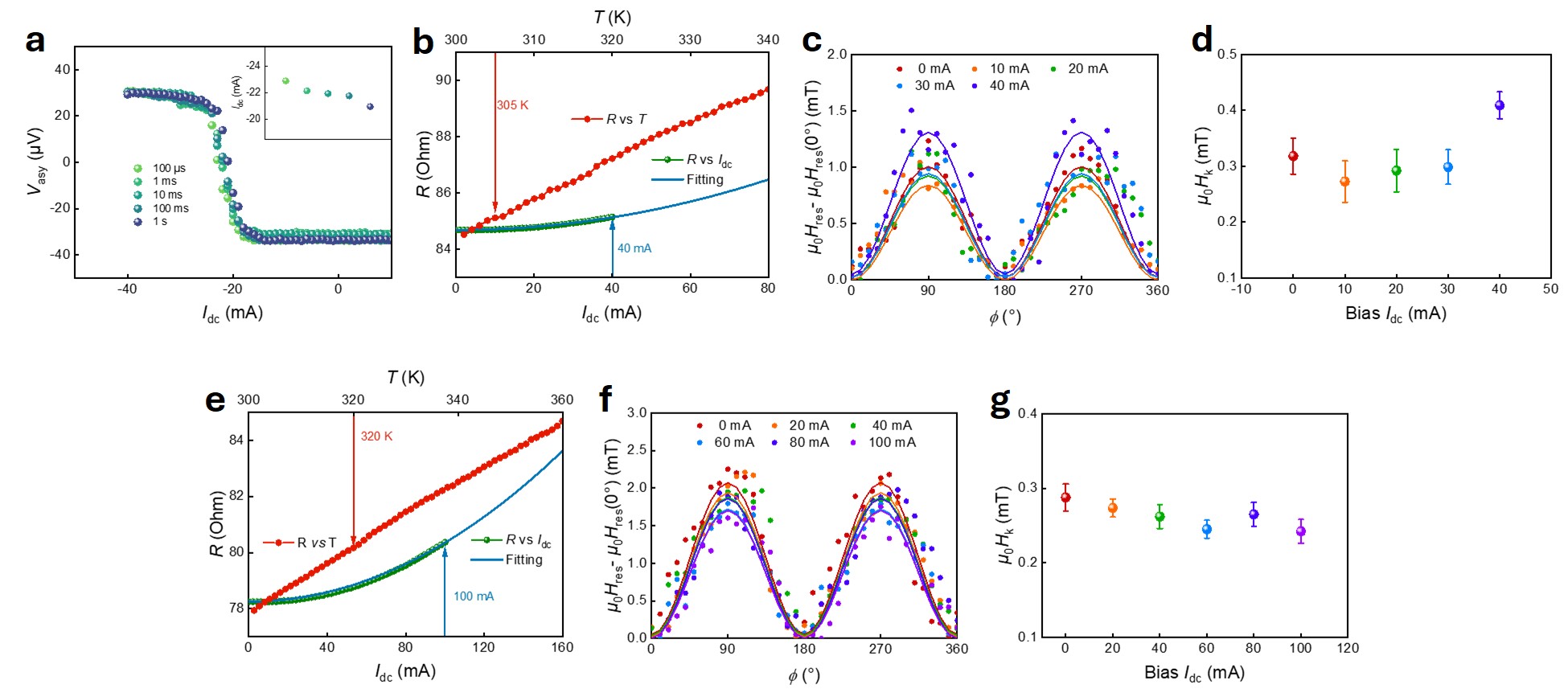}
    \caption{Pulse width and heating effect measured on device No. 9 (\textbf{a-d}) and No. 7 (\textbf{e-g}): \textbf{a}, Partial switching path driven by $I_\text{dc}$ with different pulse width probed by $V_\text{asy}$ for acoustic mode. Inset: critical switching current measured for different pulse width. \textbf{b, e} Resistance dependence on temperature ($T$) raise and Joule heating due to applied dc current ($I_\text{dc}$). \textbf{c, f} Angular dependence of resonant field ($B_\text{res}$) measured with different bias ($I_\text{dc}$) at 4 GHz (\textbf{c}) and 2 GHz (\textbf{f}), respectively. \textbf{d, g} Shape anisotropy field ($\mu_{0}H_\text{k}$) under different bias $I_\text{dc}$ extracted from fitting of \textbf{c} and \textbf{f}, respectively.}
    \label{fig:Heating}
\end{figure}

In this section, we evaluate the potential influence of Joule heating on the current-induced switching within SAF devices.

As shown in Fig.~\ref{fig:Heating} \textbf{a}, the critical switching current exhibits no dependence on the pulse width of the applied dc current varying from 1~s to \SI{100}{\micro\second}. The absence of a systematic change in the switching threshold indicates that the switching process occurs on a timescale shorter than the applied pulse durations and suggests that thermally assisted switching mechanisms make a negligible contribution to the observed behaviour.

The magnitude of current-induced Joule heating was quantified by comparing the resistance increase caused by the applied current with that resulting from an increase in ambient temperature. From the fitting of $R-I_\text{dc}$ dependence shown in Fig.~\ref{fig:Heating} \textbf{b,e}, the device resistance increases by approximately 3.3\% at an applied current of 100 mA. By comparison with the temperature-dependent resistance calibration, this resistance change corresponds to an equivalent temperature at 320 K. Although Joule heating can elevate the device temperature during operation, the resulting temperature remains well below the Curie temperature of permalloy ($\sim 850$ K), ensuring that the ferromagnetic order is preserved.

To further assess the influence of Joule heating on the magnetic properties, we quantified the shape anisotropy by fitting the angular dependence of the resonance field measured under different DC bias currents. As shown in Fig. \ref{fig:Heating} \textbf{c,d} and \textbf{f,g}, no significant variation in the extracted shape anisotropy is observed as the bias current increases from 0 to 100 mA. This result demonstrates that the magnetic anisotropy remains essentially unchanged under the current employed during the switching measurements.

In summary, these results indicate that the temperature rise induced by Joule heating remains within a regime where the magnetic properties of the SAF devices are stable. Consequently, the switching efficiency reported in this work is not significantly affected by Joule-heating effects, and the conclusions drawn from the switching measurements remain valid.

\newpage

\section*{Supplementary Note 8: Scaling of Joule heating and thermal stability with reduced dimensions}
\begin{figure}[H]
    \centering
    \includegraphics[width=0.8\linewidth]{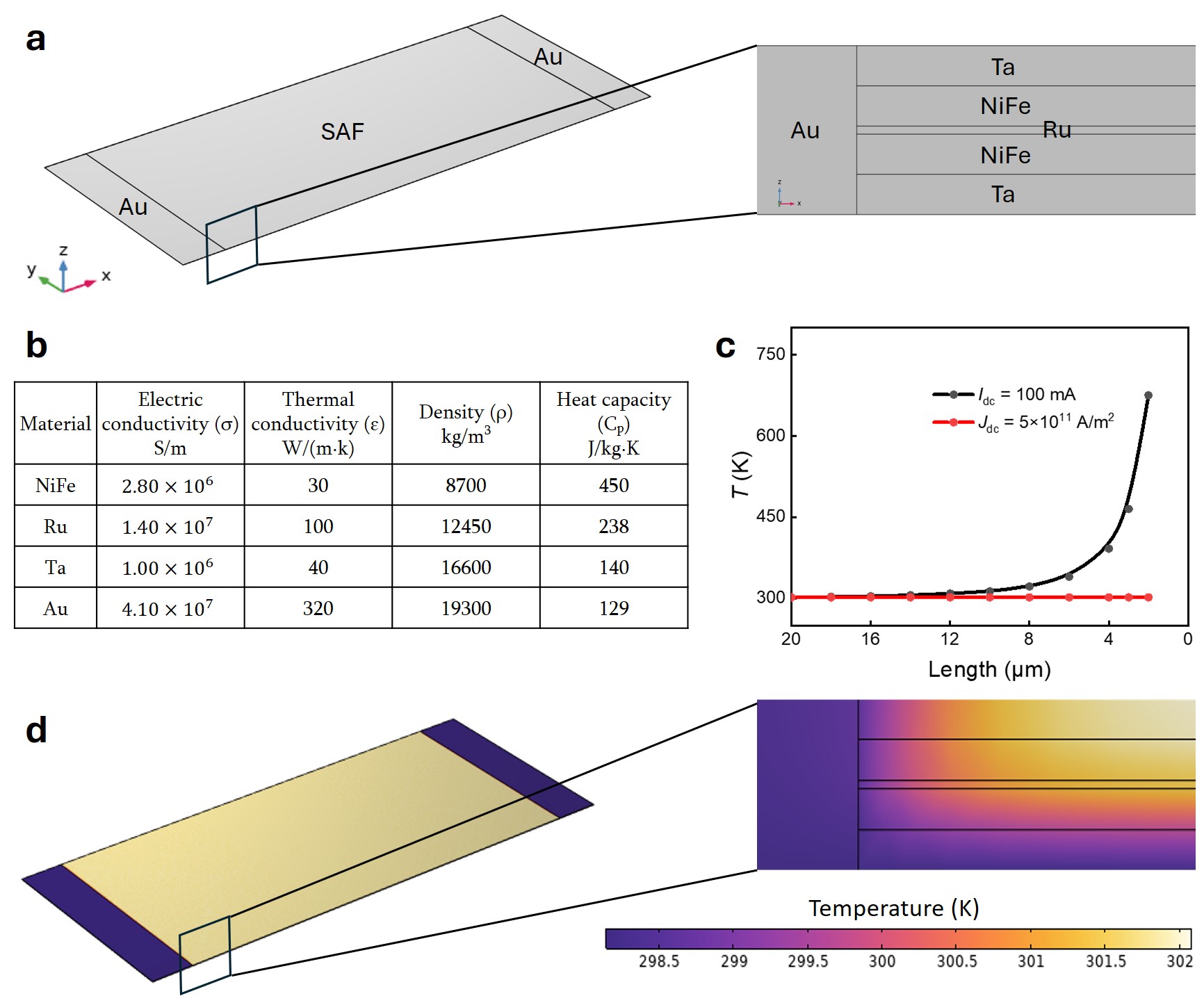}
    \caption{COMSOL simulation for heating effect within miniaturized devices: \textbf{a} Schematic of the simulated device geometry where the length/width ratio is fixed at L/W=2 so that all devices have the same resistance. \textbf{b} Table of materials parameters used for COMSOL simulation. \textbf{c} Simulated temperature rise caused by Joule heating for device lengths $L$ from 20 to 2~$\mu$m. Each device is simulated at room temperature (298~K) under either a constant current ($I_\text{dc}$=100 mA, black dots) or a constant current density ($J_\text{dc}=5\times 10^{11} \mathrm{A/m^2}$, red dots). \textbf{d} Spatial distribution of temperature under a constant current density ($J_\text{dc}=5\times 10^{11} \mathrm{A/m^2}$). }
    \label{fig:Comsol_Heating}
\end{figure}

\begin{figure}[H]
    \centering
    \includegraphics[width=0.6\linewidth]{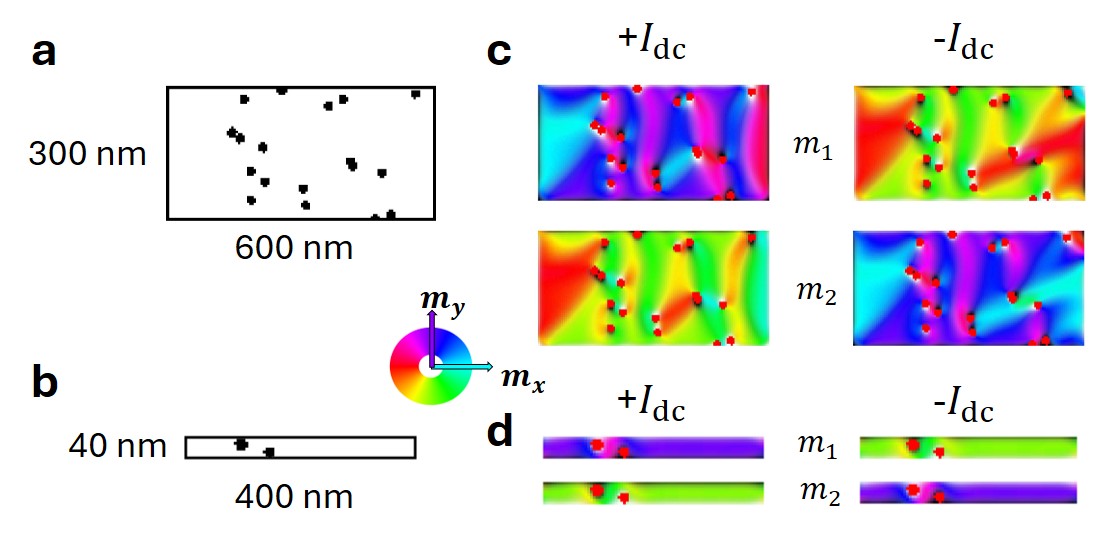}
    \caption{\textbf{a, b} Simulation of current-induced switching for smaller device dimensions: L = 600 nm, W= 300 nm for $\mu_0H_\text{k}$= 1 mT (\textbf{a}) and L = 400 nm, W= 40 nm for $\mu_0H_\text{k}$= 10 mT (\textbf{b}). The black dots in the devices are defects. \textbf{c, d} Simulated spatial mapping of magnetic moments for two layers controlled by $I_\text{dc}$.}
    \label{fig:small_device}
\end{figure}

Here we examine how switching current, current density, and thermal load scale with reduced device dimensions.

COMSOL simulations were performed to investigate the thermal distribution within our devices, as summarised in Fig.~\ref{fig:Comsol_Heating}. We miniaturised the device while keeping the length-to-width aspect ratio ($L/W$) fixed at 2, so that the device resistance is size-independent and the in-plane shape anisotropy that must be overcome for Néel-vector switching is likewise constant. The calculated device geometry is shown in Fig.~\ref{fig:Comsol_Heating}a where a mesa of a Ta/NiFe/Ru/NiFe/Ta multilayer is contacted by Au electrodes.

Using the standard material parameters as listed in Fig.~\ref{fig:Comsol_Heating}b, we simulated the device temperature $T$ as a function of $L$, taking T = 298 K at zero current (Fig.~\ref{fig:Comsol_Heating}c). Fig.~\ref{fig:Comsol_Heating}d shows the resulting spatial temperature distribution obtained under a constant current density ($J_\text{dc}=5\times 10^{11} \mathrm{A/m^2}$). For the fixed-current simulation (black), $T$ increases steeply as the device size is reduced. The resistance and the current are both constant, so the dissipated power is the same for every device size. However, this power is deposited in an increasingly small volume as $L$ shrinks, which raises $T$. For the fixed-current-density model, no such increase is observed, because the total current in a smaller device is less and the dissipated power therefore scales with the device size. As shown earlier, we have clear evidence that the Néel vector switching in our devices is governed by the current density, rather than by the current, meaning that the switching mechanism does not suffer from significant heat accumulation upon downsizing.

We extend this discussion further by considering the thermal stability of Néel vector states in synthetic antiferromagnetic systems. The two Néel vector states are separated by an energy barrier given by the shape anisotropy energy $K_\text{u}V$ where $K_\text{u}$ is the shape anisotropy and $V$ is the magnetic volume. Taking the standard thermal stability criterion: $K_\text{u}V/k_\text{B}T>60$, the minimum volume required for thermal stability is $V>60k_\text{B}T/K_\text{u}$.

Using the anisotropy energy density obtained from our experiments ($K_\text{u}=3\times10^2$ J/m$^3$) and $T=300$ K, the minimum V that satisfies this condition is $8.28\times 10^{-22}$~m$^3$. This is met, for example, by a device with $L=576$ nm, $W=288$ nm, for $t_\text{NiFe}$ = 5 nm. We therefore performed Néel vector switching simulations on such small devices as shown in Fig.~\ref{fig:small_device}. For a device with $L=600$ nm and $W=300$ nm, successful switching is observed for both polarities of $I_\text{dc}$ (Fig.~\ref{fig:small_device}a,c).

Current-induced Néel vector switching is also obtained in a smaller device with $L=400$ nm and $W=40$ nm, which has a large anisotropy ($\mu_0H_\text{k}$= 10~mT). Micromagnetic simulations were performed to show that the proposed switching and detection method can be applied for large anisotropy cases up to $\mu_0H_\text{k}$= 20~mT, as shown in Supplementary Figure 8.

These results provide promising evidence that the current-induced switching mechanism remains effective in miniaturised devices.

\newpage
\section*{Supplementary Note 9: Scanning nitrogen-vacancy magnetometry}
Scanning nitrogen-vacancy (NV) magnetometry has been performed using a commercial Qzabre Quantum Scanning Microscope system to measure the spatially resolved local magnetic fields from the SAF sample No.8. A diamond nanopillar containing a single NV centre implanted tens of nm above the base is attached to a quartz tuning fork atomic force microscopy (AFM) tip. Electrical readout of the tuning fork piezoelectric response provides an in-situ surface topography readout. The NV- axis lies along one of the four crystallographic axes belonging to the $\langle111\rangle$ family and this defines the measurement axis of the tip in use.

\begin{figure}[H]
\centering
\includegraphics[width=\textwidth]{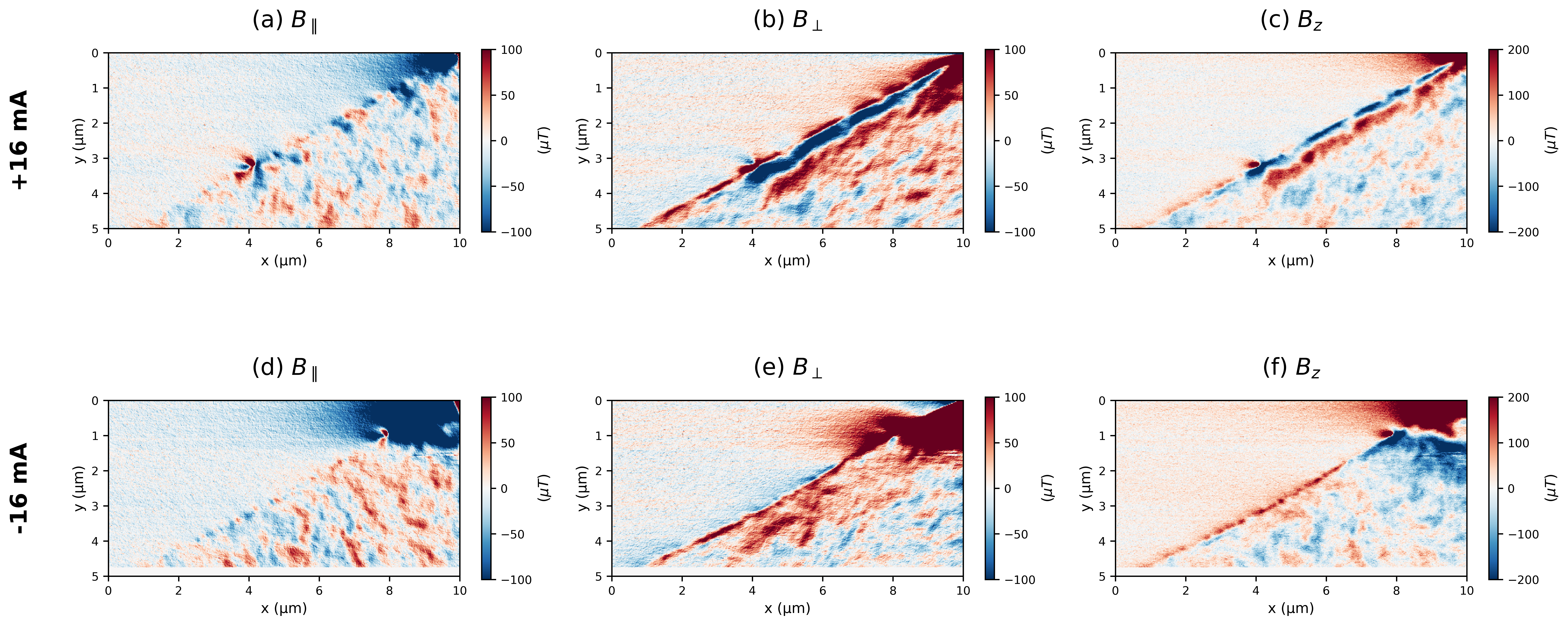}
\caption{A 2D spatial map of vector stray field components in the device edge region after an $I_\text{dc}= +16$ mA and $I_\text{dc}= -16$ mA current pulse (top and bottom row respectively). Here, $B_\parallel$ and  $B_\perp$ are defined to be parallel to  and in-plane perpendicular to the device edge respectively. The two scans are performed in the same location and any thermal drift between the scan times is corrected using cross-correlation between the two atomic force microscopy images.}
\label{fig:nv1}
\end{figure}

The NV- forms a spin-1 (triplet) defect state which allows for a spin dependent photoluminescence (PL) readout for detection of the magnetic field along the measurement axis. This was achieved using a 532 nm laser to optically initialise the $m_\text{s}= 0$ state, whilst monitoring spin dependent PL under microwave excitation slightly offset from the zero-field splitting of the NV’s magnetic dipole transition (2.87 GHz), optically detected magnetic resonance (ODMR). A small bias field $B_\text{Bias}$ ($\sim2.5$ mT along the NV measurement axis) is applied using a vector electromagnet to split the $m_\text{s}=\pm1$ resonances while minimally perturbing the magnetic state of the device.
Continuous-wave ODMR (CW-ODMR) maps, Fig.~\ref{fig:nv1}, were collected by sweeping the MW frequency through the NV's magnetic dipole transition resonant frequency ($f_\pm$) at each pixel. $f_\pm$ is modulated in the presence of external magnetic fields ($B_\text{Sample}+B_\text{Bias}$) due to the Zeeman potential ($\hat{H}_z=h\gamma \bm{S}\cdot \bm{B}$). This allows a direct conversion of the shifts in spectral response to the magnetic field projection along the NV spin axis via:
\begin{align}
    f_\pm = D \pm\gamma B_\parallel
\end{align}
where, $D$ is the zero-field splitting ~2.87 GHz, $\gamma$ is the gyromagnetic ratio $\sim28$ GHz$\cdot$ T$^{-1}$ and $B_\text{NV}$ is the total projected field along the NV’s spins axis, defined as:
\begin{align}
    B_\text{NV} (x,y)=\hat{n}_\text{NV}\cdot B(x,y,z_\text{NV})
\end{align}

The NV measures the projection of the magnetic field along the NV axis. The full vector stray field components are then reconstructed from this scalar projection by imposing the Maxwellian constraint $\nabla \cdot \bm{B} = 0$, together with the magnetostatic field's spatial dependence above the sample - typically performed in the Fourier domain. The stray field components parallel to the surface normal (or the z-direction) and the in-plane (x,y) directions are interdependent. The z-component of the stray field $B_z$ can be determined using Fourier decomposition of the field map along the measurement axis $B_\text{NV}$:

\begin{align}
    \tilde{B}_z (k,z)=\frac{\tilde{B}_\text{NV}(k,z)}{\cos\theta-i\sin\theta\Big(\cos\phi\big(\frac{k_x}{k}\big) +\sin\phi \big(\frac{k_y}{k}\big)\Big)}
\end{align}

Here, $|k|=\sqrt{k_x^2+k_y^2  }$ and represents the in plane spatial frequency vector in partial Fourier space $(k_x,k_y,z)$, $z$ is the NV fly height from the sample of interest and $\tilde B$ represents the Fourier transform of the field, $\theta$ and $\phi$ are the polar and azimuthal angles of the NV measurement axis. With the $B_z$  identified it's sufficient to define the existing field x,y components via :
\begin{align} \label{eqn:z_to_xy}
    \tilde{B}_{(x,y)}(k,z)=\frac{ik_{(x,y)}}{k \tilde{B}_{z} (k,z)}
\end{align}
The values of $\theta$ and $\phi$ for the probe used in this study were 55$^\circ$ and 0$^\circ$ respectively.
To calculate the vector stray field at a given NV lift height, the $B_z$ component is first obtained using the expression:
\begin{align}
    B_z (k,z)=\frac{e^{-kz} (1-e^{-kt} )}{2}\left(-\frac{\nabla}{k}\right)M(k)
\end{align}
With Eq.~\eqref{eqn:z_to_xy} above used for recovering the $B_x$ and $B_y$ components.
Since the direction of the device is not aligned with the scanning direction in-plane, we rotate the basis vectors of the stray field into the device frame. We used the atomic force microscopy that corresponds to the SNVM data and measure the angle $\alpha$ between the scan frame and the device edge. This is then used to transform the in-plane stray field components by a 2D rotation:

\begin{align}
    \begin{pmatrix} B_\parallel \\ B_\perp \end{pmatrix}
=
\begin{pmatrix} \cos\alpha & -\sin\alpha \\ \sin\alpha & \cos\alpha \end{pmatrix}
\begin{pmatrix} B_x \\ B_y \end{pmatrix}
\end{align}

\begin{figure}
\centering
\includegraphics[width=\textwidth]{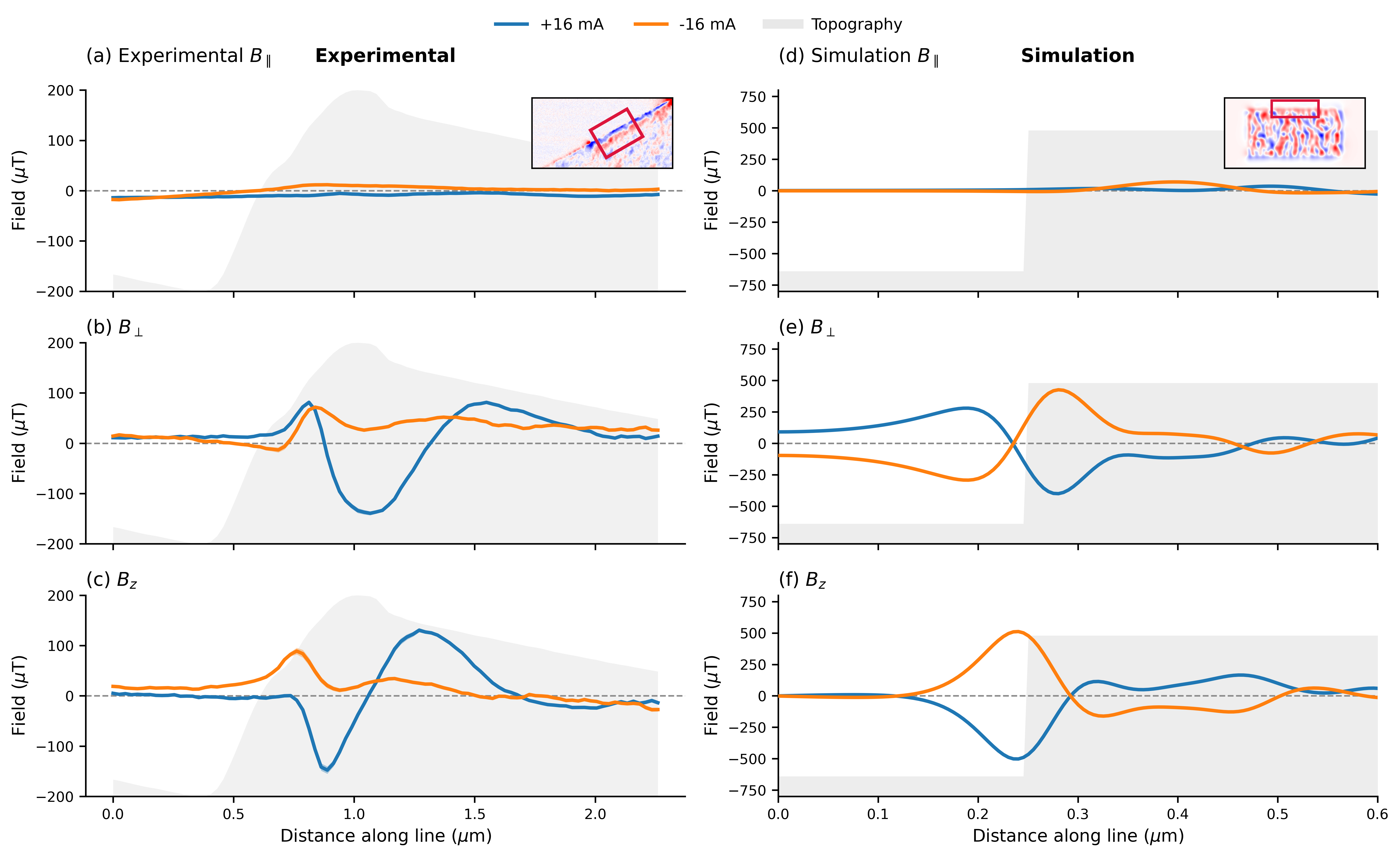}
\caption{Line profiles perpendicularly intersecting the device top edge of the stray field components in the device frame, averaged over a wide section of the edge (indicated on the inset of each columns). The stray field profiles of the relaxed magnetisation states following both the positive current and negative current cases are shown, with colour indicated by the legend at the top of the figure. The shaded background provides a normalised guide of the averaged topography of the device edge corresponding to the line profile. The two columns show the experimental data compared with simulation.  There is a general qualitative agreement in the ordering of maxima and minima between the experimental and simulation line profiles. }
\label{fig:nv2}
\end{figure}

We applied $I_\text{dc}=\pm 16$ ~mA to the device and measured the stray field projection along the $NV^-$ axis at a specific fly height, which is converted to the Cartesian vector stray-field components at the same location via Fourier space vector decomposition, with knowledge of the NV unit vector. The scanning of the same location was made possible by using one mechanical defect as a reference point to align imaging for both measurements. Figure~\ref{fig:nv1} summarises individual stray field components ($B_\perp$, $B_\parallel$ and $B_z$) after the current applications. Complex multi-domain structures are confirmed for both current polarities, and a visible difference between the two states is clearly observed, consistent with our picture of the multi-domain nature that plays a key role in current-induced Néel vector switching. To further investigate the effect of $I_\text{dc}$ on the magnetic states, we focus on the field component at the device edges that provide a robust experimental observable that can distinguish between the two magnetisation states. Next we select the edge region (indicated by the squares in Fig.~\ref{fig:nv2}) and average the perpendicular line profiles along the longitudinal direction to the device and plot the field components as a function of the perpendicular distance. Figure~\ref{fig:nv2} summarises this analysis on $B_\perp$, $B_\parallel$ and $B_z$ for $I_\text{dc}=\pm 16$ ~mA, where grey backgrounds represent the device topography to locate the edge - the sample roughness makes it less abrupt. We also present each field component calculated by our microspin simulation results in Fig.~\ref{fig:nv2}. For these plots, the stray field from each ferromagnetic layer has been calculated at the estimated NV lift height position (25~nm) and summed to obtain the total stray field observed by the NV sensor. Simulation results clearly show a sign reversal between $I_\text{dc}=\pm 16$ mA, together with complex field oscillations. The corresponding experimental results have similar oscillatory behaviours which drastically change with the application of $I_\text{dc}$. The change is complex and not a simple sign reversal for each field component, and we speculate that this is due to additional phenomena not captured by our simulations, such as complex domain configurations that would complicate simple interpretation of magnetisation reversal. A more systematic study, both experimentally and computationally, will be therefore necessary to address this open question.

\newpage

\section*{Other Supplementary Figures}
\subsection*{Supplementary Figure 1: VSM data for synthetic antiferromagnets}
\begin{figure}[H]
    \centering
    \includegraphics[width=0.5\linewidth]{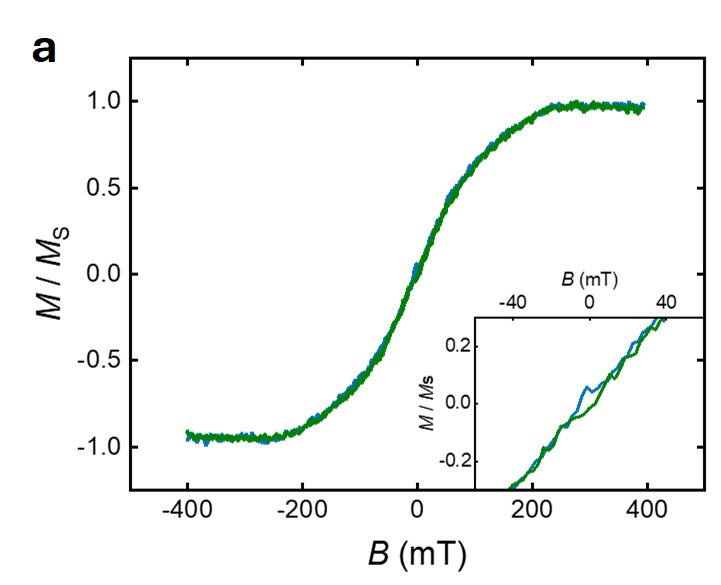}
    \caption{\textbf{a} VSM characterisation of synthetic antiferromagnet thin film with a stack of Ta(5nm)/ NiFe(5nm)/ Ru(0.3nm)/ NiFe(5nm)/ Ta(5nm). Inset: negligible remanence at zero field. }
    \label{fig:VSM}
\end{figure}

\vspace{1 cm}
\subsection*{Supplementary Figure 2: Fitting of ST-FMR spectra}
\begin{figure}[H]
    \centering
    \includegraphics[width=0.8\linewidth]{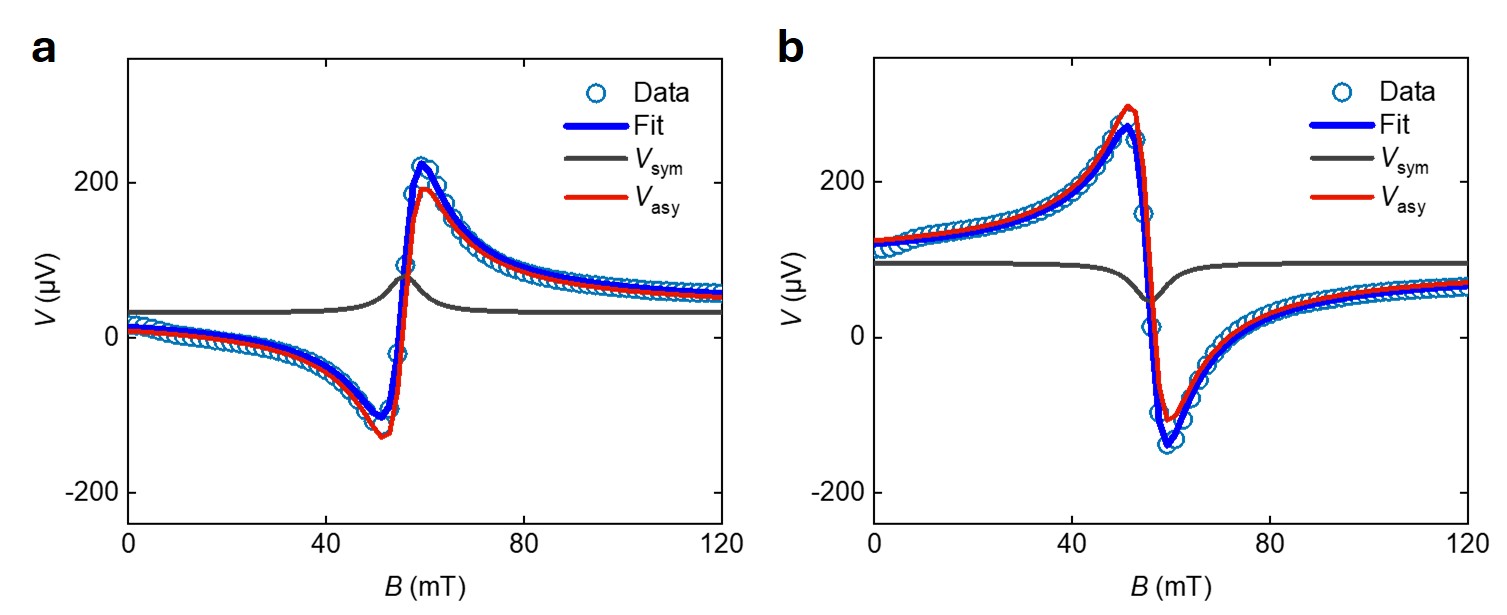}
    \caption{Fitting of ST-FMR spectra using the Lorentzian function. Rectified voltages $V$ (blue dotted line) measured in the ST-FMR scheme for device No. 8 at $I_\text{dc}=-15$ mA (\textbf{a}) and $I_\text{dc}=+15$ mA (\textbf{b}), with the fitted $V_\text{sym}$ (black solid line) and $V_\text{asy}$ (red solid line), respectively. The ST-FMR spectra are measured at 4 GHz with $B$ along $\phi$ = 60 deg. The parameters extracted from fitting of each spectrum are: \textbf{a}, $V_\text{sym}= 4.8 \times 10^{-5}$ V, $V_\text{asy}=3.2 \times 10^{-4}$ V, $B_\text{res}=55.8 $ mT, $\Delta B_\text{ac}=4.02\times10^{-3}$; \textbf{b}, $V_\text{sym}= -5.1 \times 10^{-5}$ V, $V_\text{asy}=-4.1 \times 10^{-4}$ V, $B_\text{res}=55.7 $ mT, $\Delta B_\text{ac}=3.97\times10^{-3}$.}
    \label{fig:Fitting_ac}
\end{figure}

\vspace{1 cm}
\subsection*{Supplementary Figure 3: The current-induced Néel vector reversal probed by symmetric component for acoustic mode}
\begin{figure}[H]
    \centering
    \includegraphics[width=1.0\linewidth]{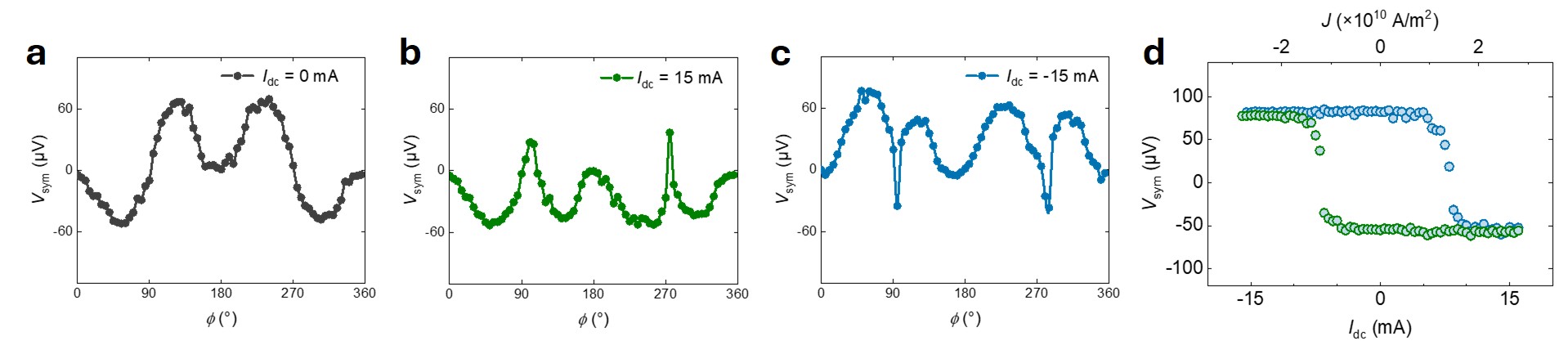}
    \caption{Current-induced switching probed by symmetric component ($V_\text{sym}$) for acoustic mode in ST-FMR: Angular dependent $V_\text{sym}$ as a function of $\phi$ for $I_{\mathrm{dc}} = 0~\text{mA}$ (\textbf{a}), $I_{\mathrm{dc}} = 15~\text{mA}$ (\textbf{b}) and $I_{\mathrm{dc}} = -15~\text{mA}$ (\textbf{c}), respectively; (\textbf{d}) $I_\text{dc}$ driven hysteresis loop measured at $\phi=60^\circ$. All results are measured on device No.8.}
    \label{fig:symmetric_components}
\end{figure}

\vspace{1 cm}
\subsection*{Supplementary Figure 4: The current-induced Néel vector reversal probed by optical mode}
\begin{figure}[H]
    \centering
    \includegraphics[width=1.0\linewidth]{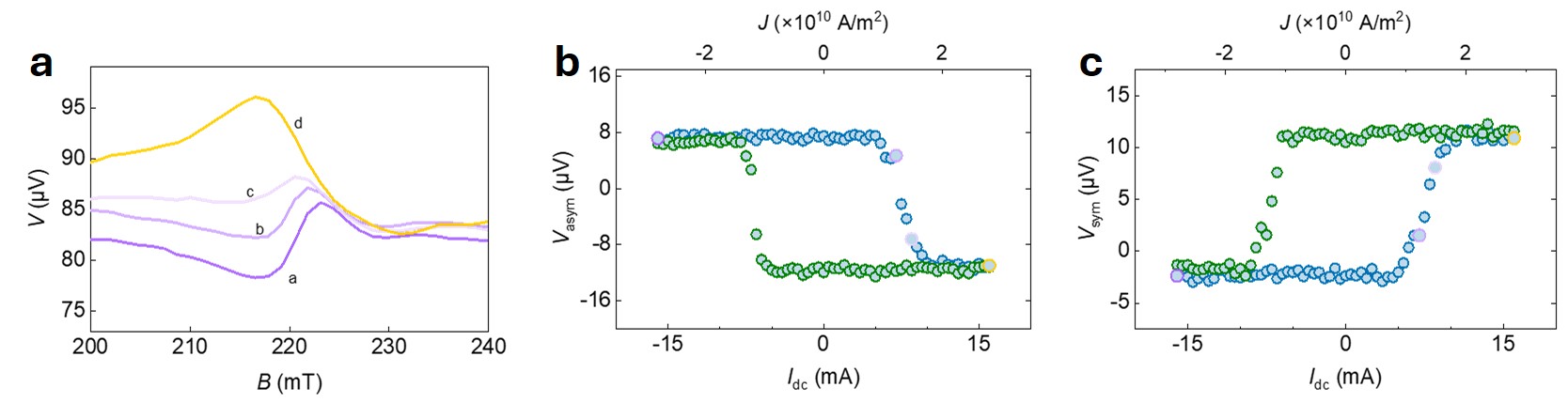}
    \caption{Current-induced switching probed by optical mode in ST-FMR: \textbf{a} Individual $V$ measurements at 4 GHz after exciting the device with different $I_\text{dc}$ (purple from deep to light for -15 mA, 7 mA and 8.5 mA, yellow for 15 mA, respectively). $I_\text{dc}$ driven hysteresis loop measured at $\phi=60^\circ$: $V_\text{asy}$ (\textbf{b}) and $V_\text{sym}$ (\textbf{c}) for optical mode. All results are measured on device No.8.}
    \label{fig:other_components}
\end{figure}

\vspace{1 cm}
\subsection*{Supplementary Figure 5: The current-induced Néel vector reversal measured in other devices}
\begin{figure}[H]
    \centering
    \includegraphics[width=0.9\linewidth]{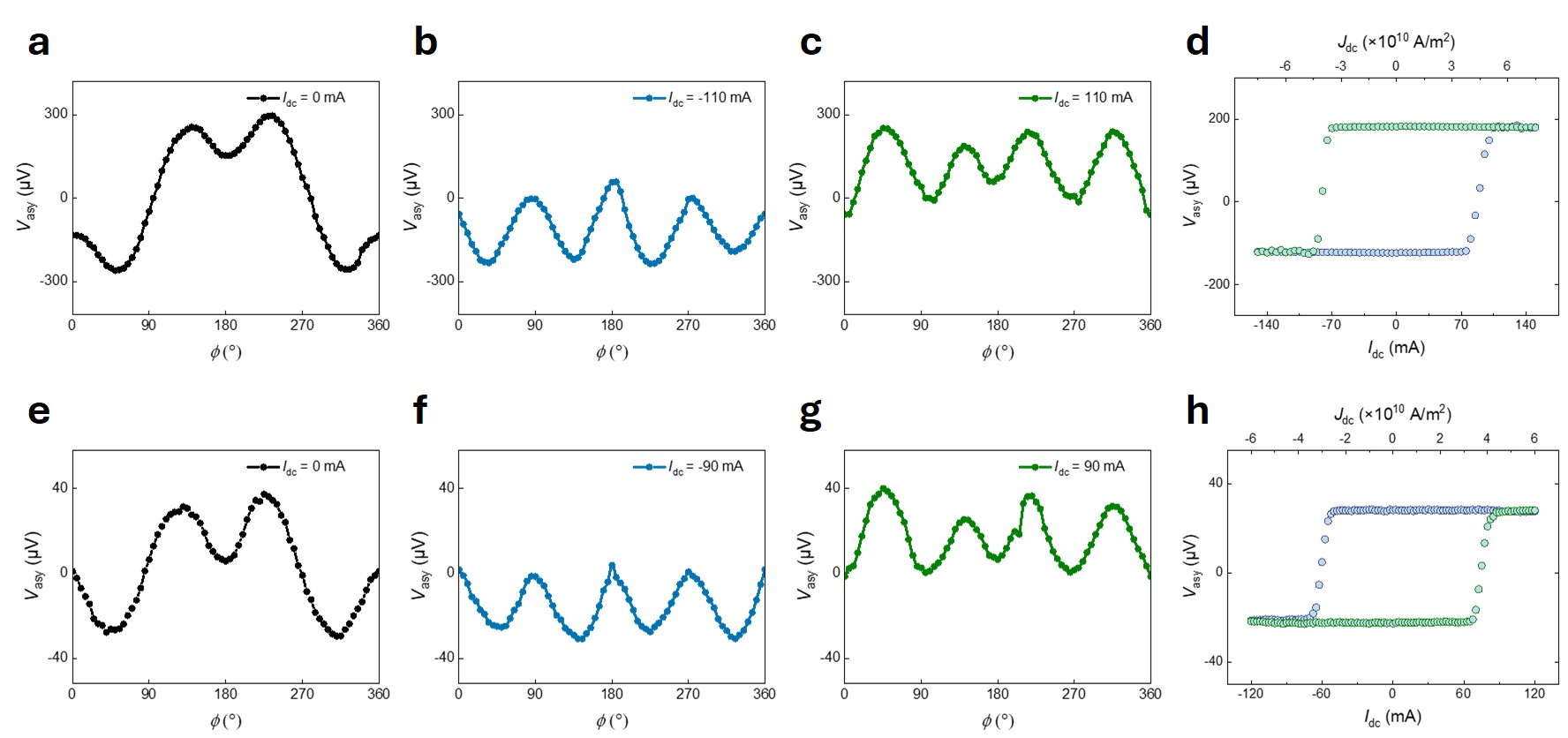}
    \caption{Spin-torque ferromagnetic resonance (ST-FMR) measurement for other devices: device No. 3 (\textbf{a-d}) and No. 5 (\textbf{e-h}). Angular dependence of $V_\text{asy}$ measured without $I_\text{dc}$ (\textbf{a, e}), with negative $I_\text{dc}$ (\textbf{b, f}) and positive $I_\text{dc}$ (\textbf{c, g}), respectively. Hysteresis loops in $V_\text{asy}$ in response to $I_\text{dc}$ are also presented for both device (\textbf{d, h}). The width (\rev{conductive} cross-sectional area) of both devices is $100\times10^{-6}~\text{m}$\ \rev{$(1.9\times10^{-12}~\text{m}^2)$}. The critical switching current density $J_\text c$ of each device are \rev{$5.53\times10^{10}~\text{A/m}^2$ and $3.42\times10^{10}~\text{A/m}^2$}, respectively, \rev{calculated assuming 1~nm oxidation of the top Ta layer}. The serial number of devices is consistent with Table~\ref{tab:all_device_profile}. }
    \label{fig:Otherdevices}
\end{figure}

\vspace{1 cm}
\subsection*{Supplementary Figure 6: Micromagnetic simulation of defect-controlled domain structures}
\begin{figure}[H]
    \centering
    \includegraphics[width=0.8\linewidth]{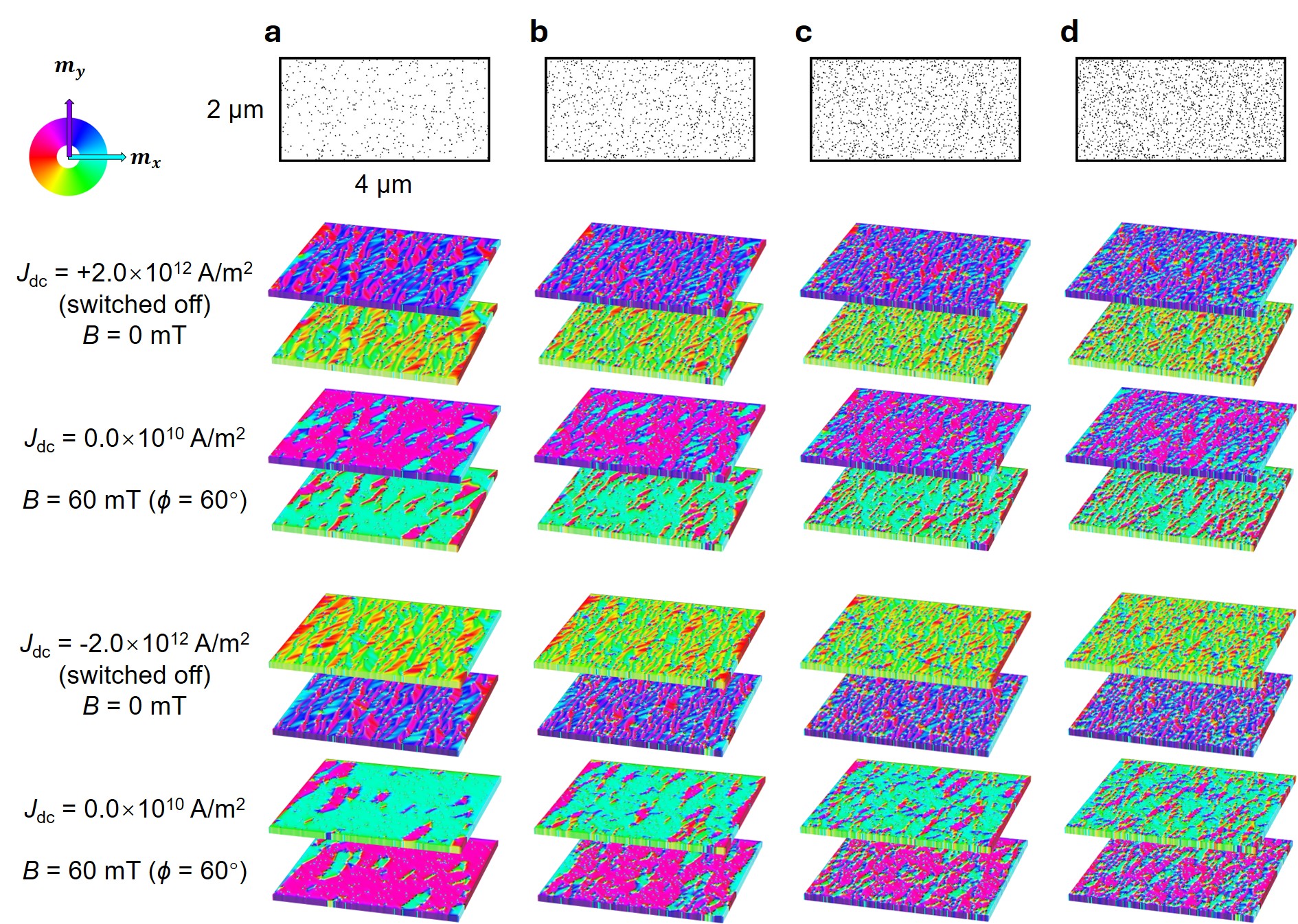}
    \caption{\textbf{a-d} Micromagnetic simulation of domain structure for systems with different defect densities: $50\times10^{12}~/\text{m}^{2}$ (\textbf{a}), $100\times10^{12}~/\text{m}^{2}$ (\textbf{b}), $150\times10^{12}~/\text{m}^{2}$ (\textbf{c}) and $200\times10^{12}~/\text{m}^{2}$ (\textbf{d}). The defect density is controlled by the number of circular vacancies that are randomly generated across the simulated region. Domain structures and magnetisation with the existence of current density ($J_\text{dc}$) or field ($B$) along certain in-plane angle ($\phi$) are presented for both magnetic layers.}
    \label{fig:mumax_defect_depend}
\end{figure}

\vspace{1 cm}
\subsection*{Supplementary Figure 7: Micromagnetic simulation of angular dependence}
\begin{figure}[H]
    \centering
    \includegraphics[width=0.8\linewidth]{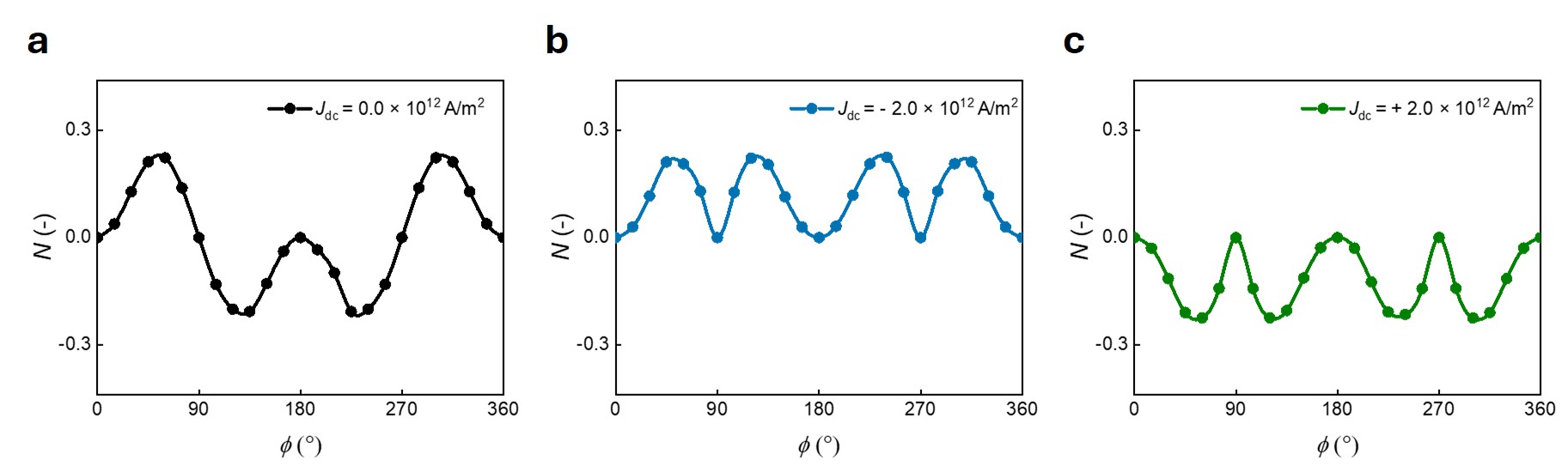}
    \caption{Micromagnetic simulation of $N$ as a function of $\phi$ for $J_{\mathrm{dc}}$ = 0 (\textbf{a}), $J_{\mathrm{dc}}$ = -2.0$\times10^{12} ~\text{A/m}^2$ (\textbf{b}) and $J_{\mathrm{dc}}$ = +2.0$\times10^{12} ~\text{A/m}^2$ (\textbf{c}). After the switch-off process of the corresponding current, magnetisation of two ferromagnetic layers ($m_1$ and $m_2$) is recorded from micromagnetic simulation where $B = 60~\text{mT}$. $V_\text{asy}$ is proportional to $N$ as indicated in Eq.~\eqref{eq:Vasy_ac} from Supplementary Note 2. }
    \label{fig:mumax_angular_simulated}
\end{figure}

\vspace{1 cm}
\subsection*{Supplementary Figure 8: Scaling of critical switching current density with higher shape anisotropy field}
\begin{figure}[H]
    \centering
    \includegraphics[width=0.7\linewidth]{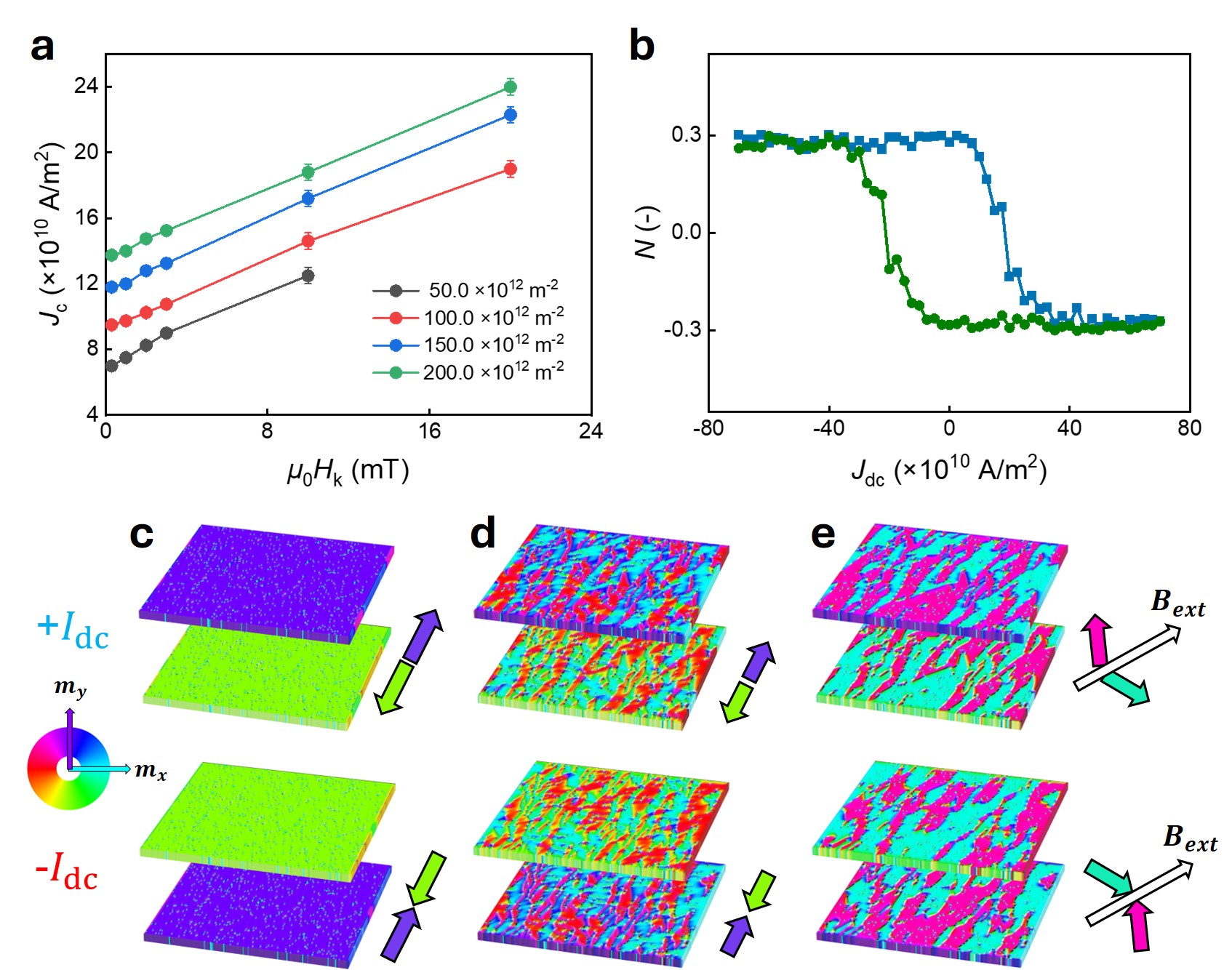}
    \caption{Néel vector switching in devices with large shape anisotropy: \textbf{a} Simulation of critical switching current density ($J_\text{c}$) scaling with shape anisotropy field ($\mu_0H_\text{k}$) from 0.3 mT to 20 mT for different defect density. \textbf{b} Simulation of Néel components ($N$) as a function of current density for $\mu_0H_\text{k}$ = 20~mT. \textbf{c-e} Simulated magnetisation components maintained during current excitation for two different current polarities (\textbf{c}), after excitation (\textbf{d}), and after application of the external magnetic field along $\phi=60^\circ$ (\textbf{e}). }
    \label{fig:larger_anisotropy_scaling}
\end{figure}

\vspace{1 cm}
\subsection*{Supplementary Figure 9: Switching efficiency in device No.2}
\begin{figure}[H]
    \centering
    \includegraphics[width=0.7\linewidth]{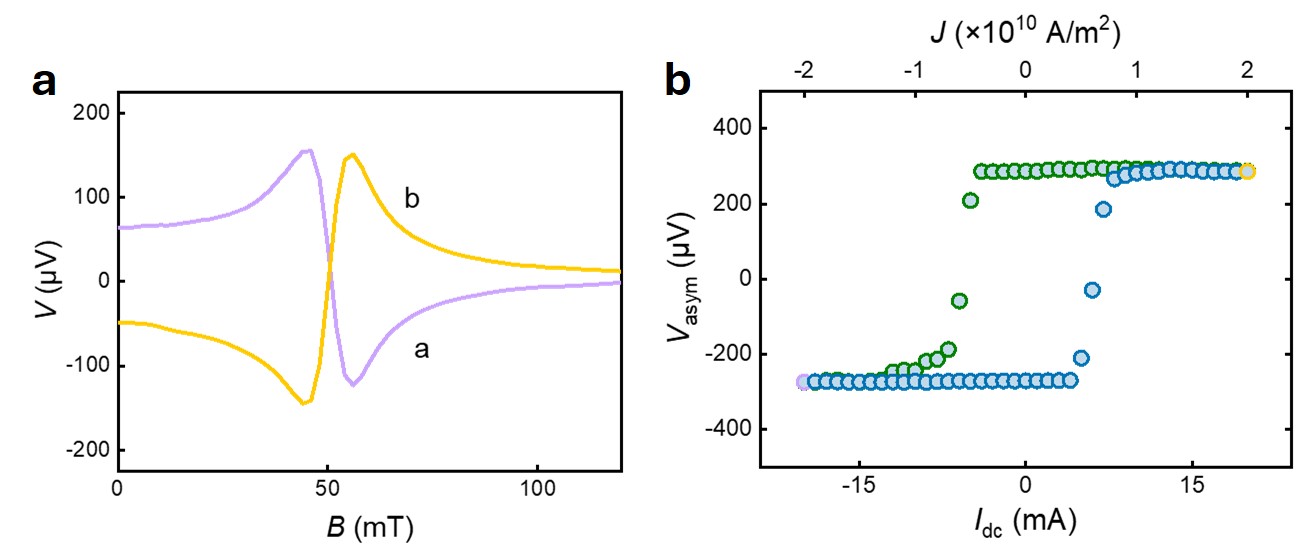}
    \caption{Current-induced switching probed by ST-FMR in device No.2. \textbf{a} Individual $V$ measurements at 3 GHz after exciting the device with different $I_\text{dc}$ (purple for -20 mA and yellow for 20 mA, respectively). \textbf{b} $I_\text{dc}$ driven hysteresis loop measured at $\phi=60^\circ$: $V_\text{asy}$  as a function of $I_\text{dc}$ for acoustic mode. The critical switching current density is \rev{$7.4\times 10^9$~A/m$^2$}, \rev{calculated assuming 1~nm oxidation of the top Ta layer, see Table~\ref{tab:all_device_profile}}. }
    \label{fig:SupFig9_device2}
\end{figure}

\vspace{1 cm}
\subsection*{Supplementary Figure 10: Non-volatility of current-induced magnetic states}
\begin{figure}[H]
    \centering
    \includegraphics[width=0.4\linewidth]{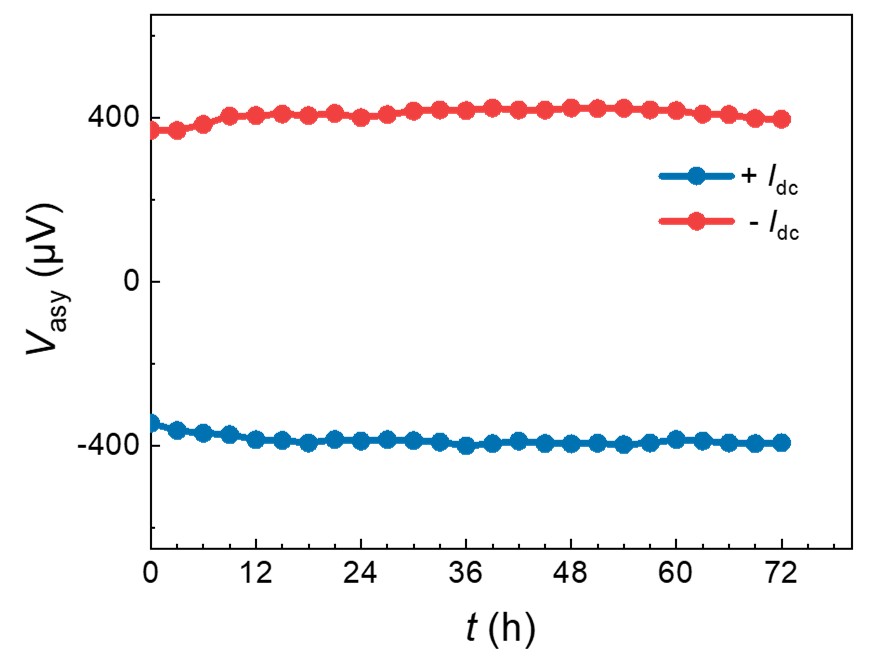}
    \caption{Time dependence of current-induced states probed by ST-FMR. A single current pulse ($\pm 15~\text{mA}$) is applied at $t = 0$, after which ST-FMR measurements are performed every 3 hours for a total duration of 72 hours. The persistence of the measured signal demonstrates that the current-induced state in our SAF device is non-volatile and exhibits stable readout over the full measurement period. }
    \label{fig:Aging}
\end{figure}

\vspace{1 cm}
\subsection*{Supplementary Figure 11: Control experiments with Cu/NiFe/Ru/NiFe/Cu multi-layers}
\begin{figure}[H]
    \centering
    \includegraphics[width=0.8\linewidth]{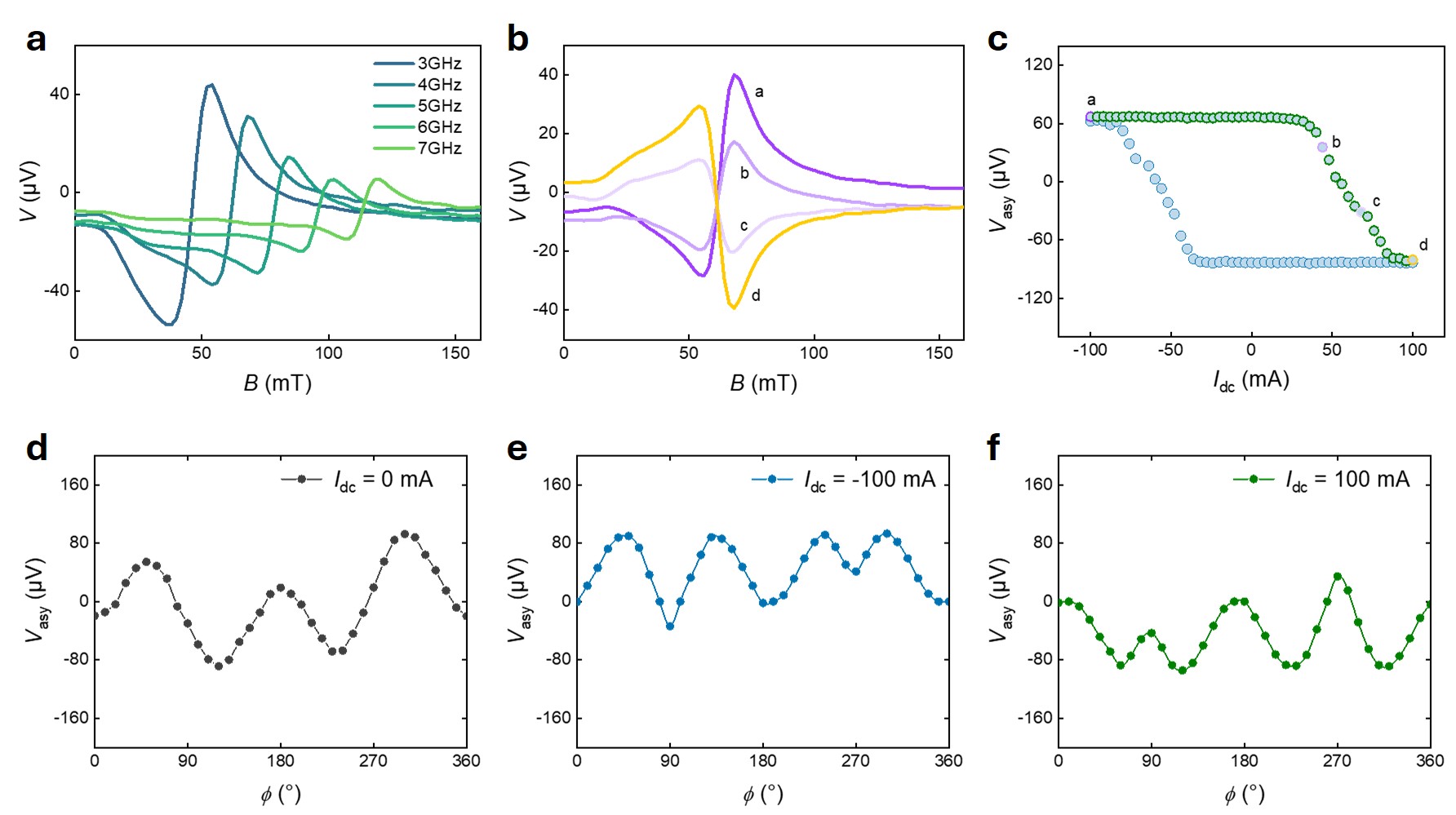}
    \caption{Control sample of Cu/NiFe/Ru/NiFe/Cu measured by ST-FMR. \textbf{a} ST-FMR spectra of control sample measured for different frequency. The antisymmetric component dominates the rectified voltage as Cu provides field-like torque more significantly than damping-like torque. \textbf{b, c} Hysteresis loop of $V_\text{asy}$ as a function of $I_\text{dc}$ measured at $\phi=45^{\circ}$. \textbf{d-f} The angular dependency of $V_\text{asy}$ measured for different $I_\text{dc}$: 0 mA (\textbf{d}), -100 mA (\textbf{e}), 100 mA (\textbf{f}). All results are measured on device No. 11. The device profile and correlating switching efficiency can be found in Table~\ref{tab:all_device_profile}. }
    \label{fig:Cu control}
\end{figure}

\end{document}